 \documentclass[pre,amsmath,amssymb,twocolumn, showpacs, superscriptaddress,10pt]{revtex4-1}
\usepackage{amsmath}
\usepackage{hyperref}
\usepackage{graphicx}
\usepackage{amsfonts}
\usepackage{amsthm}
\usepackage{cases}
\usepackage{bm}

\usepackage{comment}

\usepackage[normalem]{ulem}

\usepackage{color}
\definecolor{Blue}{rgb}{0.00, 0.00, 1.00}
\definecolor{Red}{rgb}{1.00, 0.00, 0.00}
\definecolor{Green}{rgb}{0.00, 0.70, 0.00}
\newcommand{\red}{\color{Red}}
\newcommand{\blue}{\color{Blue}}
\newcommand{\green}{\color{Green}}

\hypersetup{
    colorlinks=true,       
    linkcolor=red,          
    citecolor=blue,        
    filecolor=magenta,      
    urlcolor=cyan           
}

\newcommand{\be}{\begin{equation}}
\newcommand{\ee}{\end{equation}}
\newcommand{\bea}{\begin{eqnarray}}
\newcommand{\eea}{\end{eqnarray}}

\newcommand{\beq}{\begin{equation}}
\newcommand{\eeq}{\end{equation}}
\newcommand{\beqn}{\begin{eqnarray}}
\newcommand{\eeqn}{\end{eqnarray}}

\begin{document}

\title{The stochastic porous medium equation in one dimension}

\author{Maximilien Bernard}
\affiliation{Laboratoire de Physique de l'Ecole Normale Sup\'erieure, CNRS, ENS and PSL Universit\'e, Sorbonne Universit\'e, Universit\'e Paris Cit\'e, 24 rue Lhomond, 75005 Paris, France}
\affiliation{LPTMS, CNRS, Univ. Paris-Sud, Universit\'e Paris-Saclay, 91405 Orsay, France}
\author{Andrei A. Fedorenko}
\affiliation{ Univ Lyon, ENS de Lyon, CNRS, Laboratoire de Physique, F-69342 Lyon, France}
\author{Pierre Le Doussal}
\affiliation{Laboratoire de Physique de l'Ecole Normale Sup\'erieure, CNRS, ENS and PSL Universit\'e,
Sorbonne Universit\'e, Universit\'e Paris Cit\'e, 24 rue Lhomond, 75005 Paris, France}
\author{Alberto Rosso}
\affiliation{LPTMS, CNRS, Univ. Paris-Sud, Universit\'e Paris-Saclay, 91405 Orsay, France}
\date{\today}

\begin{abstract}
We study the porous medium equation (PME) in one space dimension in presence of additive non-conservative white noise,
and
interpreted as a stochastic growth equation for the height field of an interface.
We predict the values of the two growth exponents $\alpha$ and $\beta$ using the functional RG.
Extensive numerical simulations show agreement with the predicted values for these
exponents, however they also show {\it anomalous scaling} 
with an additional "local" exponent $\alpha_{\rm loc}$, as well as multiscaling
originating from broad distributions of local height differences.
The stationary measure of the stochastic PME is found to be well described
by a random walk model, related to a Bessel process. This model
allows for several predictions about the multiscaling properties.
\end{abstract}



\maketitle

{\bf Introduction}. The porous medium equation (PME) $\partial_t h = \nabla D(h) \nabla h$
describes 
the evolution of a conserved scalar field $h$ in a $d$ dimensional non-linear medium \cite{vazquez}. As such it appears
ubiquitously in physics, e.g. to model heat transfer in presence of non linear thermal conductivity,
or density flows of an ideal gas in porous media with non-linear equations of state \cite{barenblatt1952some,zel1950towards}.
The PME equation was proved to arise as an hydrodynamic limit of interacting particle systems
\cite{varadhan1991scaling,oelschlager1990large}, 
in kinetically constrained lattice gases \cite{goncalves2009hydrodynamic,cardoso2023derivation}.
It also arises in the conserved dynamics of magnetic interfaces \cite{spohn1993surface,spohn1993interface}
and in sandpile models \cite{bantay1992self}. 
More recently it was found to describe superdiffusion of energy in non integrable Luttinger liquids
\cite{bulchandani2020superdiffusive}. 
The PME is often studied for $D(h) \propto |h|^{s-1}$ a power law,
the case $s \to 0$ being the logarithmic PME \cite{pedron2005logarithmic}.
For general $s$ it admits special front solutions, 
known as Barenblatt's solutions \cite{barenblatt1952some,zel1950towards,vazquez}, spreading as $x \sim t^{1/(2 + d (s-1))}$
subdiffusively for $s<1$, and superdiffusively for $s>1$.


In this Letter we study a stochastic version of the porous medium equation (SPME) in presence
of additive non-conservative noise. We will consider it as a model for the
growth of an interface of height $h(x,t)$ with evolution
equation
\be \label{spme} 
\partial_t h(x,t) = \nabla^2 V(h(x,t)) + \eta(x,t) 
\ee
where $\eta(x,t)$ is a Gaussian white noise 
uncorrelated in time and space,
$V(h)$ is an increasing function of $h$, and 
the Laplacian term can also be written as $\nabla D(h) \nabla h$
with $D(h)=V'(h)$, i.e. $V(h) \propto h |h|^{s-1}$ in the power law case.
For $s=1$ it reduces to the celebrated Edwards-Wilkinson (EW) equation \cite{edwards1982surface}, 
i.e. the simplest linear equation describing the equilibrium growth of an interface. 
In that case the height fluctuations scale as $h \sim x^\alpha$ with
roughness exponent $\alpha=1/2$, while the space-time scaling
$t \sim x^z$ corresponds to dynamical exponent $z=2$. 
For $s \neq 1$
the non-linearities make the problem non-equilibrium, as for the Kardar-Parisi-Zhang equation \cite{kardar1986dynamic},
although here the universality class is different since the equation is conservative. 



An anisotropic version of the SPME in space dimension $d=2$ has been studied before in the context of erosion
of tilted landscapes \cite{pastor1998stochastic,pastor1998scalingJStat,antonov2017scaling,birnir2013mathematical,pelletier1999self}
and of continuous models for self-organized criticality \cite{hwa1989dissipative}. 
More recently the stochastic PME has also been studied in the context of the Wilson-Cowan equation
in critical neural networks models of the brain dynamics
\cite{tiberi2022gell} as well as in active matter
\cite{speckactive}. Furthermore the stochastic PME has also been studied in mathematics \cite{barbu2016stochastic}
where existence of solutions was obtained both
for the additive noise \cite{da2004weak,kim2006stochastic}
and multiplicative noise
\cite{barbu2009stochastic,dareiotis2021porous}.


Despite these works, there is no quantitative theory for the SPME in $d=1$, unlike the KPZ equation. The aim of the present work is thus to determine the 
scaling exponents $\alpha$ and $z$, and to numerically check whether standard scaling applies to
this highly non-linear stochastic growth equation \cite{nagatani}.
Surprisingly, we find that the scaling for $s \neq 1$ is {\it anomalous}, suggesting  the presence of a second roughness exponent $\alpha_\text{loc}$. Although  {\it anomalous} scaling has been observed in other systems \cite{lopez1999scaling,krug1994turbulent, Randomdiff,lopez1998anomalous}, its origin is fundamentally different in the context of SPME.

\begin{figure}
\centering
\includegraphics[width=0.48\textwidth]{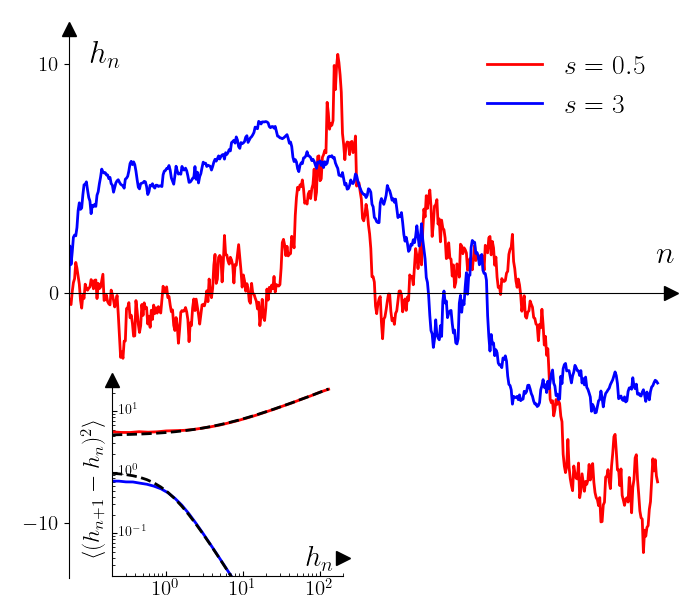}
\caption{Examples of stationary interfaces for  $s=3$, $\nu=1$ (blue) and $s=0.5$, $\nu=0.25$ (red) $L=500$ . The latter has been rescaled by a factor $5$ for aesthetic purposes. For $|h_n|$ away from zero, the interface becomes rougher when $s<1$ and flatter when $s>1$.
In the inset, the mean square increment of the interface height, $\langle (h_{n+1}-h_n)^2 \rangle$ at fixed value of $h_n$, is compared to the mean square jumps $1/D(h_n)$ of the random walk defined in Eq. \eqref{eq:RW} (dashed lines) \cite{SM}. We observe a perfect agreement at large $h_n$.
}
    \label{fig:figint}
\end{figure}

In particular, we consider the following discrete version of the SPME in $d=1$ with additive white noise:

\be
\label{eq:SPMEdx}
\partial_t h_n = D(h_{n-1}) \left[h_{n-1}-h_{n} \right] + D(h_n) \left[h_{n+1}-h_{n} \right] +\eta_n(t).
\ee
Note that in the above equation, the time dependence of $h_n$ is kept implicit. This equation represents the time evolution of a spring chain characterized by non-homogeneous and $h_n$-dependent elastic constants $D(h_n)$. Different behavior of $D(h_n)$ can be considered with the only constraint that $D(h_n) > 0$. In this paper, we study the power-law case, $D(h_n) \sim |h_n|^{s-1}$ for large $h_n$ with $s>0$ and $D(h_n)$ even. At the origin, the behavior is regularized with $D(0)=\nu$ (See \cite{SM} for the precise form of $D(h)$ used in numerical simulation). In  Fig.\ref{fig:figint}, we show two interfaces in the stationary regime. The interface with $s>1$ is  flatter compared to the case $s<1$. Indeed, for large $h$, since $D(h) \sim |h|^{s-1}$ the springs become stiff when $s>1$ and soft when $s<1$.
One of the main result of this paper is 
the following prediction for the roughness and dynamical exponents
\be \label{RGexponentsavecd} 
\alpha = \frac{2-d}{1+ s} \quad , \quad z = 2 - \frac{(2-d) (s-1)}{s+1} 
\ee 
This prediction verifies the exact exponent relation\cite{wolf1990growth}
\be 
z = 2 \alpha + d,
\ee 
valid for conservative dynamics with non conserved
noise \footnote{Indeed, integrating the SPME equation \eqref{spme} 
over the system volume leads to $h \sim t^{1/2}/L^{d/2}$, and thus to the exponent relation.}.
 First, we perform a functional RG calculation and find stable fixed points that preserve the same power law behavior $D(h) \sim  |h|^{s-1}$ at large $h$. 
This leads to the prediction Eq. \eqref{RGexponentsavecd} for $d \leq 2$, obtained
to first order in a $\epsilon=2-d$ expansion, but argued to be exact for $s>0$. 
However, we also find that local observables exhibit {\it anomalous scaling}.  
For instance, for $\ell \ll L,$ and $n \sim L$, the $q$ moments of the {\it stationary} equal time height-to-height correlation read:
\be 
\label{eq:anomscalstatq}
 \langle |h_{n+\ell} - h_n|^q \rangle^{1/q} \sim 
 \ell^{\alpha_{\rm loc}(q)} L^{ (\alpha - \alpha_{\rm loc}(q))}
\ee 
where $\langle \dots \rangle$ indicates the average over the noise. 
In Eq.\eqref{eq:anomscalstatq}, $\alpha_{\rm loc}(q)$ is 
commonly referred to as the {\it local} roughness exponent \cite{lopez1999scaling,krug1994turbulent, Randomdiff,lopez1998anomalous} and can depend on the value of $q$. 
The anomalous scaling arises due to the presence of inhomogeneous increments, $h_{n+1} - h_n$, along the interface.


 Third, in the final part of the Letter, we show with heuristic arguments and numerical simulations, that the stationary interface of the SPME, provided that $L$ is large, has the same statistical properties as the following random walk:


\be 
\tilde{h}_{n+1} = \tilde{h}_n + \frac{\chi_n}{\sqrt{D(\tilde{h}_n)}} 
\label{eq:RW}
\ee
where $\chi_n$ is a centered unit Gaussian random variable, independent of $\tilde{h}_n$. Using this mapping, we can recover the roughness exponent from Eq.\eqref{RGexponentsavecd} as well as the anomalous scaling. 
In particular, we predict that $\alpha_\text{loc}(q)=1/2$ for $0<s<1$ while for $s>1$, one has 

\begin{align} \label{eq:alphalocqs1}
 &
 \alpha_\text{loc}(q) \sim  \begin{cases} 1/2 ~~~~   \quad ,\quad  q < q_c \\
 \alpha + \frac{1}{q} \frac{s}{1+s} \quad ,\quad  q_c < q 
 \end{cases} 
\end{align}
with $q<q_c=2+2/(s-1)$.

{\bf Renormalization group}. To perform the RG analysis of the SPME \eqref{spme} we use the Martin-Siggia-Rose dynamical field theory in space dimension $d$,
near the upper critical dimension $d=2$, in a perturbative expansion in $\epsilon=2-d$. 
The RG analysis of this model was first performed in \cite{antonov1995quantum} within a polynomial expansion of the function $V(h)$ keeping only first few terms. Here we analyze it in the full functional form, that is necessary to identify all possible fixed points which are determined by the large $h$ behavior of $V(h)$.
We introduce an infrared renormalization scale $\mu$ (an inverse
length scale which can be taken as $L^{-1}$). We denote the bare potential in the dynamical action by $V_0(h)$,
where $V'_0(0)=\nu$ is the bare diffusivity,
and the 
$\mu$-dependent renormalized potential at scale $\mu$ by $V(h)$, defined from the effective action. The 
functional RG flow to one loop (i.e. to first order in $\epsilon>0$) then reads \cite{SM} 
\be \label{aa}
- \mu \partial_\mu V(h)  = \frac{\mu^{-\epsilon}}{4 \pi} \frac{V''(h)}{V'(h)},
\ee
We introduce 
$u=\frac{1}{4 \pi} \int_{\mu}^{\mu_0} d\mu'/(\mu')^{1+\epsilon}$ so that $u \sim \mu^{-\epsilon}/\epsilon$ for $\epsilon>0$
and $u= \frac{1}{4 \pi} \log(\mu_0/\mu)$ in $d=2$. 
Let us denote the running $V'(h)$ at the scale $\mu$ by $D_u(h)$. The RG equation can be put in the form
\be \label{RG2main} 
\partial_u D_u(h) = \partial_h^2 \log D_u(h)
\ee
which, interestingly, is itself a logarithmic PME.
We can look for scale invariant fixed point
solutions of the RG flow in the form
\be \label{eqn:scaleWmain}
D_u(h) = u^{1- 2a} F( u^{-a} h),
\ee 
Once we find such a fixed point solution, 
since $u \sim \mu^{-\epsilon} \sim L^\epsilon$, one can identify $\alpha = a \epsilon$ where $\alpha$ is the anomalous scaling dimension of $h \sim L^\alpha$.
The corresponding time scale is $t \sim L^2/\nu(L)$ where $\nu(L) = D_u(0) \sim L^{\epsilon(1-2 a)}$ is the
renormalized diffusivity, leading to $t \sim L^z$ with $z=2 - \epsilon(1- 2 a)$. Hence 
the exact relation $z=2 \alpha+d$
follows and there is a single exponent (as the noise is not corrected
under RG) see \cite{SM} for details.
Inserting \eqref{eqn:scaleWmain} into \eqref{RG2main} we obtain 
the following fixed point equation for $F(H)$ with $H=h/u^a$
\be
\label{eqn:F}
a H F'(H) - (1-2a) F(H) + \frac{F''(H)}{F(H)} - \frac{F'(H)^2}{F(H)^2} = 0.
\ee




Since here we consider only the case where $V'(h)$  is finite at 
$h=0$ and even, we must have $F'(0)=0$.
Moreover, if $F(H)$ is solution of \eqref{eqn:F}, so is $b^2 F(H b)$ for any $b$. We 
therefore choose $F(0)=1$.
We have found through analytical and numerical analysis that there are four types of such solutions to the equation \eqref{eqn:F}:

(i) a line of fixed points with $0<a<1$ and $F(H) \sim_{H \gg 1}  
H^{-2+1/a}$.

(ii) a special solution $F(H) = 1/(1+ H^2/2)$ associated to $a=1$.

(iii) $F(H) \sim_{H \gg 1} e^{-C_1 H}/H^2$ associated to $a>1$.

(iv)  solutions diverging  at finite $H=H_0$ as $\sim 1/(H_0-H)$ 
for $a < 0$ and $\sim 1/(H_0-H)^2$ for $a=0$ .

\begin{figure}  \label{fig:flowz}
    \includegraphics[width=0.48\textwidth]{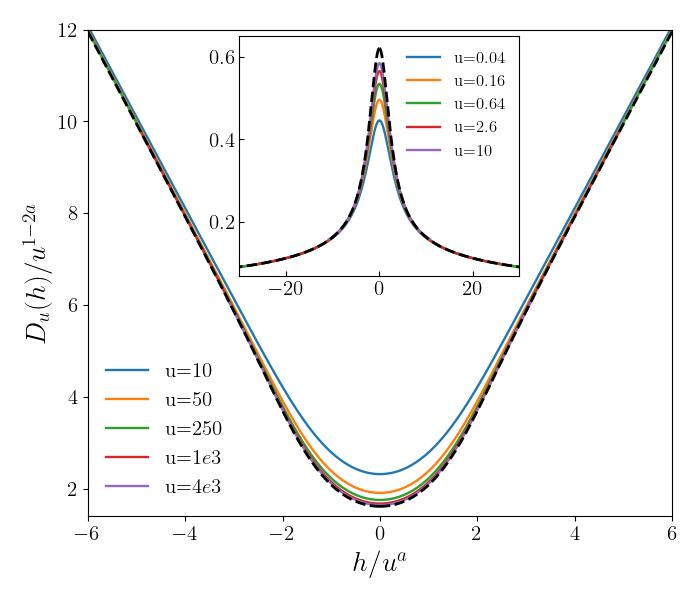}
    \caption{Numerical solution of Eq.\eqref{RG2main} illustrating the collapse of $D_u(h)$. Main panel, $s>1$: Starting from $D_0(h) =2 \sqrt{4 +h^2}$ which corresponds to $s=2$ and $a=1/(1+s)=1/3$. As $u$ increases, the solutions converge towards a self-similar fixed point, shown by the black dashed lines, which corresponds to $F(H)$ (with $b^2=1.612$), calculated numerically from \eqref{eqn:F}. 
    Inset, $s<1$: Starting from $D_0(h) = \frac{1}{2}(1/4^4+h^2)^{-1/4}$
    corresponding to $s=0.5$ and $a=2/3$. The dotted line correspond to the predicted fixed point (with $b^2=0.62$).
    }
    \label{fig:figRG}
\end{figure}

We show by linear stability analysis, see \cite{SM}, that the solutions (i) which are relevant for this paper
correspond to attractive fixed points of the RG flow Eq.\eqref{RG2main}. 
In order to study in more details the basin of attraction of these fixed points, we solve Eq.\eqref{RG2main} with several initial conditions (See Fig.\ref{fig:figRG}).
For $s>0$, the large $|h|$ behavior of $D_u(h) \sim |h|^{s-1}$ is conserved by the RG flow, which converges towards the FP (i) with $a=1/(1+s)$.
Finally, one can argue that the higher order loop corrections to the RG equations \eqref{aa} and 
\eqref{eqn:F} cannot change the large $h$ behavior of $V(h)$ for $s>0$, leading to our conjecture that the prediction \eqref{RGexponentsavecd} for the
exponents is exact. 

\begin{figure} 
    \includegraphics[width=0.49\textwidth]{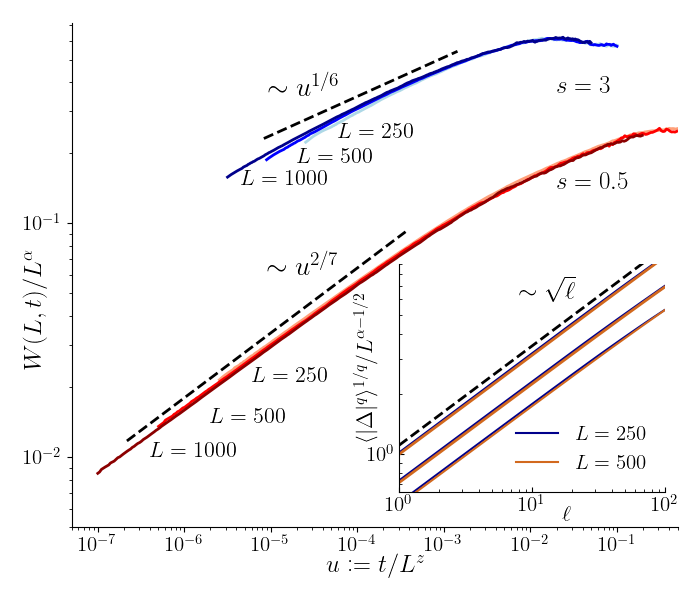}
    \caption{Main Panel: Family-Vicksek collapse \eqref{Family-Vicsek} for $d=1$ interfaces with $s = 0.5$ (continuous red line) and $s = 3$ (blue line) for several system sizes $L$. The exponents $\alpha$ and $z$ are those obtained from the RG \eqref{RGexponentsavecd}, leading to a successful collapse of the width onto a single curve.
    The black dotted line corresponds to the predicted short time growth $\sim u^\beta$, with $\beta = \alpha/z = 1/(3+s)$. The data for $s=3$ has been shifted to the left by a factor $10$ to improve the data presentation. Inset: Moments $\langle |\Delta|^q \rangle$ of the increments $\Delta = h_{n+\ell}-h_n$ for $1 \le n \le L-\ell$ and $s=0.5$. Here, $q=1,2,4$ from bottom to top. We obtain a succesful collapse using the scaling proposed in \eqref{eq:anomscalstatq} with $\alpha_{\rm loc} = 1/2$.
    }
    \label{fig:collw}
\end{figure}

{\bf Numerical results}. We use an Euler scheme to integrate Eq.\eqref{eq:SPMEdx}, from a flat initial condition, see \cite{SM} for details and for our choice
of function $D(h)$. We fix $h_0=0$ and let the right boundary free ($D(h_L)=0$).
We first compute the time evolution of the width $ W(L,t) = \sqrt{\frac{1}{L} \langle \sum_{n=1}^L (h_n(t) - \bar h(t))^2 \rangle}$,
where $\bar h(t)= \frac{1}{L} \sum_{n=1}^L h_n(t)$ denotes the center of mass. Fig.\ref{fig:collw} shows a very good agreement with Family-Vicsek scaling:
\be \label{Family-Vicsek}
W(L,t) = L^\alpha f_w( t L^{-z}).
\ee 
Here the values of the exponents $\alpha$ and $z$ are the values predicted by the RG (Eq.\eqref{RGexponentsavecd}).
The scaling function $f_w(u)$ depends on the boundary conditions, with $f_w(u) \to {\rm cst}$ for $u \gg 1$ and $f_w(u) \sim u^{\alpha/z}$ for $u \ll 1$. Indeed,  the interface is rough up to a length $\xi(t) \sim t^{1/z}$, and is flat above this scale. As a consequence, at short times ($u \ll 1$), the width $W(L,t)$, as well as the typical height $h_n(t)$, grow as $\sim \xi(t)^\alpha \sim t^\beta= t^{\alpha/z}$. 

The width is a global observable which probes the behavior of the entire interface. We have also computed local observables, particularly the $q$-th moments of the \textit{stationary} height-to-height equal-time correlation. In the inset of Fig.~\ref{fig:collw}, we show, for $s=1/2$, that it exhibits \textit{anomalous scaling} behavior for $\ell \ll L$, as displayed in Eq.\eqref{eq:anomscalstatq},
with the exponent $\alpha_{\rm loc}(q) = 1/2$. Here, the average is taken over both the noise but also over $n \lesssim L$. Although Eq.\eqref{eq:anomscalstatq} holds for the stationary state, the {\it anomalous scaling} can be extended to finite times by replacing $L$ by $\xi(t)$. We will argue that $\alpha_{\rm loc}(q) = 1/2$ for $s < 1$, while for $s > 1$ it 
depends explicitly on $q$ (See Eq.\eqref{eq:alphalocqs1}). 
Finally, we have also studied the structure factor of the height field which also displays 
the anomalous scaling \cite{SM}. 

\begin{figure}
\centering
\includegraphics[width=0.48\textwidth]{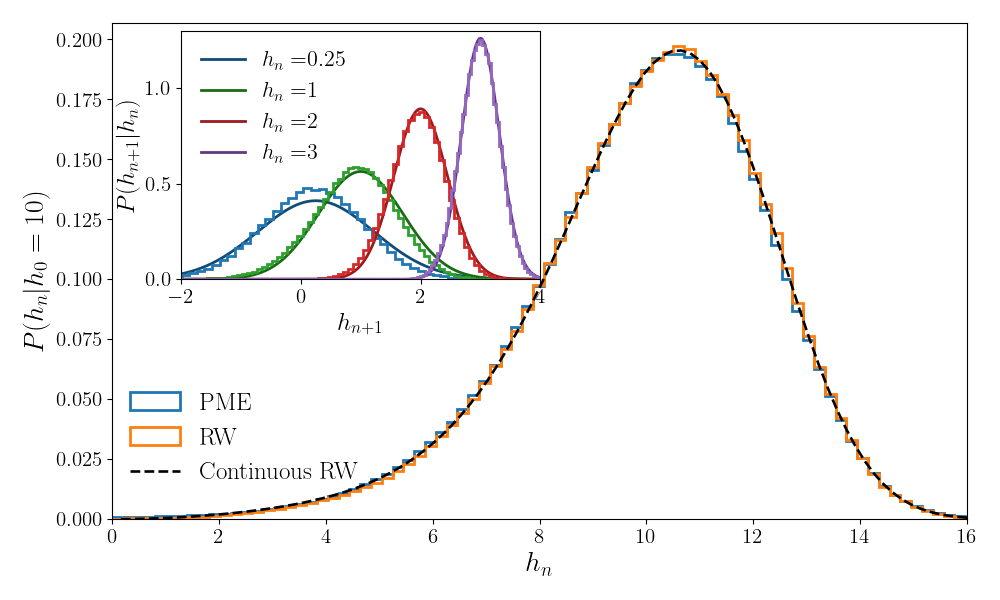}
\caption{ PDF of $h_n$ averaged over $n \in [L-p,L]$ with $L=500$ and $p=50$, with boundary condition $h_0=10$
and $D(h)=1+h^2$ ($s=3$) for (i) the stationary SPME (orange), (ii) the discrete RW model (blue),
(iii) the analytical prediction (black) from the continuum diffusion model (using the
two point transition probability in 
\eqref{eq:propagatorAbsh}) (note that there is no fitting parameter). Inset: Distribution of the height $h_{n+1}$ conditioned to a given $h_n$ for stationary interfaces of the SPME (Histogram). It is compared to the prediction for the random walk defined in Eq.\eqref{eq:RW}, which is a Gaussian of mean $h_n$ and variance $1/D(h_n)$ (Continuous lines)
}
    \label{fig:compRWPME}
\end{figure}

{\bf Connection to the random walk.} To clarify the relation between the random walk (RW) model defined in Eq.\eqref{eq:RW} and the stationary interfaces of the SPME, we recall that if the spring constants were independent of $h$, i.e. for the EW model, the stationary limit would be described by the equipartition theorem. This theorem predicts Gaussian increments with zero mean and variance $1/D$. This reasoning can be extended to cases where $D(h_n)$ varies slowly along the $x$-axis, namely $D(h_n) \simeq D(h_{n+1})$. Using a Taylor expansion, we can rewrite the latter condition as a condition for the increments: 

\be
\label{eq:condgauss1}
|h_{n+1} - h_n| \ll \frac{D(h_n)}{|D'(h_n)|} \sim  |h_n|,
\ee
where the asymptotic behavior $\sim |h_n|$ holds only for large $h_n$.
For small increments, the distribution of $h_{n+1}$ conditioned to $h_n$ becomes Gaussian of mean $h_n$ and variance $1/D(h_n)$. For large increments, deviations from the Gaussian behavior are observed. However, as can be seen in the inset of Fig.\ref{fig:compRWPME}, 
these deviations disappear when $|h_n| \gg 1$, in which case the condition \eqref{eq:condgauss1} is satisfied
\footnote{The condition Eq.\eqref{eq:condgauss1} for the increments can be rewritten as a condition for $h_n$. Indeed, for small increments, we expect $|h_{n+1} - h_n| \sim 1/\sqrt{D(h_n)}$, hence from Eq.\eqref{eq:condgauss1}, the condition becomes $|h_{n+1} - h_n| \sim 1/\sqrt{D(h_n)} \sim |h_n|^{(1-s)/2} \ll D(h_n)/|D'(h_n)| \sim |h_n|$. This condition is satisfied for $|h_n| \gg 1$.}.
This is also confirmed in the inset of Fig.\ref{fig:figint}, where we show that $\langle (h_{n+1}-h_n)^2 \rangle = 1/D(h_n)$ for large $|h_n|$.
In the stationary regime, for $n = O(L)$, the height of the interface scale as $h_n \sim L^\alpha \gg 1$, hence, in the limit of large system size, we may expect that the statistical properties of the SPME are described by the random walk model of Eq.\eqref{eq:RW} at all scales.
This is indeed what we observe, as detailed in \cite{SM} and illustrated in Fig. \ref{fig:compRWPME}. 

This RW model allows for several predictions that we now detail. 
Since the $\xi_n$ are i.i.d one obtains the second moment of the height difference (removing the tilde
symbol)
\be 
\langle (h_{n+\ell} - h_n)^2 \rangle = \sum_{m=n}^{n+\ell-1} \langle \frac{1}{D(h_m)} \rangle 
\ee 
At large $n$, we assume the scaling $h_n \sim n^\alpha {\sf h}$, where $\alpha$ is the Hurst exponent and ${\sf h}$ is the rescaled height, a random variable independent of $n$. Using $D(h) \simeq c |h|^{s-1}$ we obtain for $n \gg 1$ (see \cite{SM} for details) 
\be  \label{eq:scalehdiffrw}
 \langle (h_{n+\ell} - h_n)^2 \rangle  \simeq \frac{1}{c} \langle   {\sf h}^{1-s}\rangle \sum_{p=0}^{\ell-1} (n+p)^{\alpha (1-s)} 
\ee   
To determine $\alpha$ we consider $\ell \sim n$, in which case the r.h.s behaves as $n^{\alpha(1-s)+1}$ while
the l.h.s behaves as $n^{2 \alpha}$. Equating the two gives $\alpha=1/(1+s)$ for the RW model, 
which coincides with the prediction \eqref{RGexponentsavecd} for the SPME in $d=1$.
Furthermore, one can study the local behavior for $\ell \ll n$. 
From \eqref{eq:scalehdiffrw} we obtain
\begin{equation}
\langle (h_{n+\ell}- h_n)^2 \rangle
\sim \left\{
    \begin{array}{ll}
        n^{2\alpha} & \mbox{if } \ell \sim n \\
        n^{2\alpha-1} \ell & \mbox{if } \ell \ll n
    \end{array}
\right.
\end{equation}
which shows that the RW model exhibits anomalous scaling with $\alpha_\text{loc}(q=2)=1/2$.

To obtain all the moments of the height difference $\Delta= h_{n+\ell} - h_n$ 
we need the distribution of the rescaled height at large $n$.
This distribution can be obtained using the continuum version of the RW model, analyzed in \cite{SM}.
There we show that this continuum model is related to a Bessel process in the
variable $Y= |h|^{(s+1)/2}$. It allows to compute the distribution of the height difference, for given $n$ and $\ell \ll n$ \cite{SM}.
Its expression depends on the value of $s$. For $s>1$, it behaves as
\begin{align} \label{eq:distsp1}
 &
 P_{n,\ell}(\Delta) \sim  \begin{cases} n^{\frac{1}{2}-\alpha} \ell^{-\frac{1}{2}}~~~~~~~~~~~~~~~~~   ~ ,~  |\Delta| \ll \Delta_\text{typ}\\
 n^{-\frac{s}{s+1}} \ell^{\frac{s}{s-1}} |\Delta|^{-(3+\frac{2}{s-1})}  ~ ,~ \Delta_\text{typ} \ll |\Delta| \ll \Delta_c 
 \end{cases} 
\end{align}
with a typical value $\Delta_\text{typ}=n^{\alpha-1/2} \ell^{1/2}$ and an upper cutoff $\Delta_c=\ell^\alpha$. From \eqref{eq:distsp1}, we can calculate the $q$-th moment, which shows a multiscaling behavior
\begin{align} \label{eq:distdeltasp1mt}
 &
 \langle |\Delta|^q \rangle \sim  \begin{cases} \ell^{\frac{q}{2}} n^{q \, (\alpha-\frac{1}{2})}    \quad ,\quad  q < q_c \\
 \ell^{q \alpha + \frac{s}{s+1}} n^{-\frac{s}{s+1}}  \quad ,\quad  q > q_c 
 \end{cases}
\end{align}
with $q_c= 2+\frac{2}{s-1}$. Indeed, for $q<q_c$, the average is dominated by the typical values, namely $ \sim \Delta_{\rm typ}^q$, while for $q>q_c$, it is dominated by the upper cutoff $\sim \Delta_c^{q+1} P_{n,\ell} (\Delta_c)$. 
For $s<1$, the power law decay of $P_{n,\ell}(\Delta)$ for $\Delta_{\rm typ} \ll \Delta \ll \Delta_c$ is replaced by a stretched exponential decay. As a consequence the moments do not display multiscaling and behave as 
\begin{align}
 &
 \langle |\Delta|^q \rangle \sim  \ell^{\frac{q}{2}} n^{q \, (\alpha-\frac{1}{2})}.
\end{align}
Averaging over $1 \le n \lesssim L$, we obtain $\alpha_{\rm loc}(q)=1/2$ for $s<1$, while, for $s>1$, we recover the multifractal $\alpha_{\rm loc}(q)$ given in \eqref{eq:alphalocqs1}.
The continuum model enables further predictions, including the determination of the one-point distribution of the rescaled height $\sf h$
(see also \cite{fa2003power,giuseppe}) 
\be  
P({\sf h}) \sim |{\sf h}|^{s-1} e^{-\frac{2 c |{\sf h}|^{s+1}}{(s+1)^2}}.  
\ee  

For $0 < s < 1$, this distribution exhibits an integrable divergence at the origin. Conversely, for $s > 1$, the distribution vanishes at the origin and develops a bimodal shape. Furthermore, the two-point height probability, derived in Eq.~\eqref{eq:propagatorAbsh}, is depicted in Fig.~\ref{fig:compRWPME} and shows excellent agreement with the SPME. This mapping thus provides a detailed characterization of the statistical properties of the stationary interface.

{\bf Conclusion}. In summary, we have studied the scaling behavior of the stochastic porous medium equation with a diffusion coefficient $D(h) \sim |h|^{s-1}$ for $s>0$,
and obtained the critical exponents.
We have predicted and verified numerically that it obeys anomalous scaling with
non-trivial multiscaling properties for $s>1$. We have found that the stationary interfaces are remarkably well described by a random walk model, which enables many analytical predictions. Its large scale continuum limit is related
to a Bessel process, allowing for calculation of, e.g., the stationary height distribution as well as the multifractal exponents.
Using functional RG method we have shown that for arbitrary even functions $D(h)$ there exists only four families of fixed points finite at $h=0$. While only one of them is relevant for the scaling behavior of the  SPME, the others could also have applications. In particular, our functional RG equation can be used to
study the models of brain dynamics, such as considered in \cite{tiberi2022gell}.
{\it Acknowledgments:} 
We thank Bjorn Birnir for discussions.
AAF acknowledges support from the ANR Grant No. ANR-18-CE40-0033 (DIMERS),
AR from the ANR Grant No. ANR-23-CE30-0031-04 and PLD from the
ANR Grant ANR-23-CE30-0020-01 (EDIPS).
This work was performed using HPC resources from GENCI-IDRIS (Grant 2024-AD011015488).

\bibliography{biblio}

\newpage
.
\newpage

\begin{widetext} 

\setcounter{secnumdepth}{2}

%
\begin{center}
	\textbf{\large  The stochastic porous medium equation in one dimension
		\\
		[.3cm] -- Supplemental Material --} \\
	[.4cm] Maximilien Bernard,$^{1,2}$ Andrei A. Fedorenko,$^3$ Pierre Le Doussal,$^1$ and Alberto Rosso$^2$\\[.1cm]
	{\itshape $^1$Laboratoire de Physique de l'Ecole Normale Sup\'erieure, \\ 
       CNRS, ENS and PSL Universit\'e, Sorbonne Universit\'e, \\ Universit\'e Paris Cit\'e, 24 rue Lhomond, 75005 Paris, France} \\
{\itshape $^2$LPTMS, CNRS, Univ. Paris-Sud, Universit\'e Paris-Saclay, 91405 Orsay, France} \\
	{\itshape $^3$Univ Lyon, ENS de Lyon, CNRS, Laboratoire de Physique, F-69342 Lyon, France}\\
\end{center}

\bigskip
In this Supplemental Material, we give the principal details of the calculations described in the main text of the Letter.
\bigskip

\tableofcontents

\renewcommand{\theequation}{S\arabic{equation}}
\setcounter{equation}{0}

\section{Functional RG approach}


\subsection{Dynamical action and field theory} 
%

We study the isotropic SPME equation in space dimension $d$ 
\be \label{eqmo1}
\partial_t h(x,t) =  \nabla_x^2  V_0(h(x,t)) + \eta(x,t), 
\ee
where $\eta(x,t)$ is a Gaussian white noise with zero mean and variance
\be
\left\langle \eta(x,t) \eta(x',t')\right\rangle  = 2\sigma_0 \delta(x-x')\delta(t-t').
\ee
Here we use the subscript zero to denote the bare quantities.
We use the dynamical action, obtained by enforcing
the equation of motion using an auxiliary response field $\hat h=- i \tilde h$, and averaging over the noise.
As a result the average of any observable
${\cal O}[h_s]$ of the solution $h=h_s$ of \eqref{eqmo1} over the noise can be written as a path integral
\be
\langle {\cal O}[h_s] \rangle = \int {\cal D} h {\cal D} \tilde h \, {\cal O}[h] \, e^{- S[\tilde h, h] }
\ee
where the dynamical action reads
\be \label{dynaction}
S[\tilde h, h] = \int dt d^d x  [ \tilde h (\partial_t h - \nabla_x^2 V_0(h))  - \sigma_0 \tilde h^2 ]
\ee
and one may include the initial conditions in the definition of the path integral if needed. An infrared cutoff $\mu$
is implicit everywhere to regularize the system at large distance (alternatively one may use a finite system size $L \sim 1/\mu$).
Splitting $V_0(h)=\nu_0 h + U_0(h)$, where $\nu_0$ is the bare diffusivity  and $U_0(h)$ can be treated as the non-linear interaction, one sees from the quadratic part of the action
that the bare dimensions of the fields are $h=L^{(2-d)/2}$ and $\tilde h=L^{- (d+2)/2}$, with scaling $t \sim x^{z_0}$ with $z_0=2$, and that
the non-linear part $U_0(h)$ becomes relevant for $d \leq 2$. One thus expects logarithmic divergences in
perturbation theory at the upper critical dimension $d=d_c=2$. Since the field $h$ is dimensionless
there one expects the need for functional RG to describe the flow of the full function $V_0(h)$.

One proceeds by computing the effective action functional $\Gamma[\phi] = \Gamma[\tilde h,h]$ associated
to the action $S[\phi] = S[\tilde h, h]$, using the two component field notation $\phi=(\tilde h,h)$
whenever convenient. Let us recall that one first defines
the partition sum in presence of sources, $Z[j]= \int {\cal D}\phi e^{- S[\phi] + \int dt d^d x j(x) \cdot \phi(x)}$.
Then $W[j]=\log Z[j]$ is the generating function of the connected correlations of the field (obtained by taking successive derivatives w.r.t $j$).
$\Gamma[\phi]$ is then defined as the Legendre transform of $W[j]$,
i.e. $\Gamma[\phi]= \int dt d^d x j(x) \cdot \phi(x) - W[j]|_{W'[j]=\phi}$.
Its important property is that
connected correlations of $\phi$ are now tree diagrams obtained from $\Gamma[\phi]$. Hence
the vertices of $\Gamma[\phi]$ are the dressed (or renormalized) vertices which sum up all loops.
Decomposing $S=S_0+S_{\rm int}$ where $S_0$ is the part of the action which is quadratic in the fields,
it can be computed from the perturbative formula
\be \label{eq:loop-expansion}
\Gamma[\phi]= S_0[\phi] + \sum_{p \geq 1} \frac{(-1)^{p+1}}{p!} \langle \left( S_{\rm int}[\phi + \delta \phi] \right)^p \rangle^{1PI}_{\delta \phi, S_0}
\ee
where $\langle \dots \rangle^{1PI}_{\delta \phi, S_0}$ denote an average over the field $\delta \phi$ (at fixed background field $\phi$)
keeping only 1-particle irreducible diagrams (i.e. connected diagrams in the vertices $S_{\rm int}$ which
cannot be disconnected by cutting a propagator from $S_0$). The RG analysis then proceeds by
considering the leading dependence of $\Gamma[\phi]$ in the IR cutoff $\mu$ as $\mu \to 0$,
which is related to the analysis of the divergences in the 1PI diagrams (loop diagrams).

\subsection{Effective action and RG flow}

The calculation of $\Gamma[\phi]$ associated with the action \eqref{dynaction}, the analysis of the divergences, and the ensuing RG analysis was 
pioneered in \cite{antonov1995quantum}. Here, we thus only sketch the main ideas, and refer the interested reader to that paper for
details, as well as to the subsequent paper~\cite{antonov2017scaling}, which discusses an anisotropic version of the model studied in the context of erosion.

There it was found that to one loop it takes the same form as the action \eqref{dynaction}
\be
\Gamma[\tilde h, h] = \int dt d^d x  [ \tilde h (\partial_t h - \nabla_x^2 V(h))  - \sigma \tilde h^2 ] + \dots
\ee
up to terms which are irrelevant near $d=2$ (higher gradients). Indeed, one can check that the term $\tilde h \partial_t h$ is not corrected,
and that, furthermore, the term $\tilde h^2$ is also not corrected, so that ${ \sigma=\sigma_0}$ (and we choose ${ \sigma_0}=1$ below).
It is easy to see since a
shift $\tilde h \to \tilde h + c$ by a uniform constant $c$ in \eqref{dynaction} produces only quadratic terms,
up to boundary terms.
Equivalently, the term $ \int d^d x \tilde h  \nabla_x^2 V_0(h) $ can be integrated by part into
$\int  d^d x  V_0(h)   \nabla_x^2 \tilde h$ hence no correction to the uniform part of $\tilde h$ can
be produced. As a consequence, no wave-function renormalization, i.e. no renormalization of the field $h$, is needed
and the only quantity which flows is $V(h)$, the $\mu$-dependent renormalized potential.

In  \cite{antonov1995quantum} it was shown that the divergent part of the one loop contribution to the effective action reads,
near $d=2$, with $d=2-\epsilon$ 
\be \label{eq:Gamma1}
 \tilde \Gamma_1(\tilde h,h) =  - 
  A_d \frac{\mu^{-\epsilon}}{\epsilon} \int dt d^d x
\frac{U''(h)}{\nu + U'(h)} \nabla_x^2 \tilde h,
\ee
where $\mu$ is the IR regulator (renormalization mass) {and $A_2=1/(4 \pi)$. Note that in \cite{antonov1995quantum} the one loop correction \eqref{eq:Gamma1} was used to study the RG flow only for the first few coefficients in the Taylor expansion of $V(h)$, see below. However, in order to identify all possible fixed points, it is necessary to consider the full functional flow of $V(h)$, which is done in this work. To that end we compute
the one-loop correction $\delta V(h)$ to the full potential $V(h)$ expressed in terms of the functional diagram
\be
\delta V(h) = \includegraphics[width=6mm]{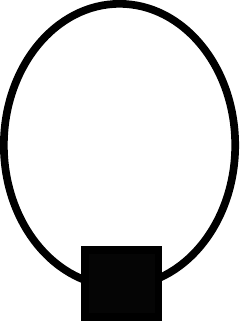},
\ee
corresponding to the lowest order correction in \eqref{eq:loop-expansion}.  Here the black square stands for the vertex $V_0''(h)$ and the solid line for the
correlation function of $\delta h$ in the presence of the background field $h$,
\be
\langle \delta h_{k,\omega} \delta h_{k',\omega'} \rangle =
\frac{2}{ D_0^2(h)(k^2+\mu^2)^2+\omega^2} \times (2 \pi)^{d+1} \delta^{(d)} (k+k') \delta(\omega+\omega')  ,
\ee
where we include the IR cutoff $\mu$. Evaluating the integrals and expanding in $\epsilon=2-d$ we arrive at
\be
\delta V(h) = \frac12 \int \frac{d\omega}{2\pi} \frac{d^d k }{(2\pi)^d}
\frac{2V_0''(h)}{ D_0^2(h)(k^2+\mu^2)^2+\omega^2}
=  A_d \frac{\mu^{-\epsilon}}{\epsilon}
\frac{V_0''(h)}{V_0'(h)} + O(\epsilon^0).
\ee
}
Taking a derivative w.r.t. $\mu$ at fixed $V_0$, and then replacing $V_0 \to V$, we obtain
the RG flow for the full potential $V(h)$ to leading order in $\epsilon$ as
\bea \label{aa2}
- \mu \partial_\mu V(h)  = \frac{1}{4 \pi} \mu^{-\epsilon} \frac{V''(h)}{V'(h)}.
\eea
This equation can alternatively be obtained from the exact RG functional equation for $\Gamma[\phi]$.
The general exact RG method was developed in \cite{wetterich1993exact,morris1994exact} and
leads to an exact RG equation for the functional $\Gamma[\phi]$ (see also Section 3 in \cite{balents2005thermal}
for a simplified summary).
To arrive at \eqref{aa2} requires
truncations, which can be performed in a systematic way in a multilocal expansion \cite{schehr2003exact,le2010exact},
and within a controlled dimensional expansion near $d=2$. These truncations can be performed more
generally within the non-perturbative FRG approach, which was applied to the anisotropic version of the model (erosion)
in
\cite{duclut2017nonuniversality,DuclutThesis}. Here we consider perturbative FRG {applied to} the isotropic model.
\\

{\bf Comparison with the RG flow of \cite{antonov1995quantum}.} Let us now compare our functional
RG flow \eqref{aa2} to the one displayed in \cite{antonov1995quantum}. To this aim we define a scaled potential $\hat V$ via
\bea \label{Vhat}
V(h) = \mu^{-\epsilon/2} \nu^{1/2} \hat V(H= \nu^{1/2} \mu^{\epsilon/2} h)  \quad , \quad \hat V(H) = H + \sum_{n \geq 2} \frac{g_n}{n!} H^n
\eea
where the $g_n$ are the dimensionless couplings defined in \cite{antonov1995quantum}. Note that $V,\tilde V,\nu,g_n$ implicitly depend on $\nu$.
By definition one has $V'(0)=\nu$ and $\tilde V'(0)=1$. Taking derivatives at $h=0$ of the RG flow equation \eqref{aa2}, and using
\eqref{Vhat},
we first obtain the flow of the diffusivity $\nu$ as
\be \label{diff}
- \mu \partial_\mu \log \nu = \frac{1}{4 \pi}  (g_3-g_2^2)
\ee
which coincides (23a) in \cite{antonov1995quantum}.
Using \eqref{diff} and \eqref{Vhat}, taking into account that $\nu$ also depends on $\mu$ via \eqref{diff} we
obtain
\bea
&& - \mu \partial_\mu g_2
= \frac{1}{2}  g_2 \epsilon + \frac{1}{8 \pi}  ( 2 g_4 - 9 g_2 g_3 + 7 g_2^3) \\
&& - \mu \partial_\mu g_3=  \epsilon g_3 + \frac{1}{8 \pi}  ( 2 g_5 - 8 g_2 g_4 - 10 g_3^2 + 28 g_2^2 g_3
- 12 g_2^4)
\eea
which is in agreement with (23b) in \cite{antonov1995quantum} (their $\beta_n$ functions have
by definition the opposite sign and there is an obvious sign misprint in the first term of their $\beta_2$).


It was argued in \cite{antonov1995quantum} that  the infinite-dimensional space
of the couplings $g_n$  has a two dimensional surface of fixed points $\{g_n^*\}$ parameterized by the values of $g_2^*$ and $g_3^*$,
and the conclusion was that studying the nature of these fixed points is a difficult task. The authors of \cite{antonov1995quantum} summarize: "The general conclusion is pessimistic: the problem of building a satisfactory microtheory of the given phenomenon remains unresolved."  In fact this problem can not be resolved  once one truncates  the $\beta_n$ as done in \cite{antonov1995quantum}. Indeed, as we show below, the fixed point is selected by the large $h$ behavior
of the bare $V_0(h)$ and to identify it one needs to consider the full functional flow.

\subsection{Critical exponents} The critical exponents are associated to scale invariant solutions of the RG flow.
Let us first examine the form that these solutions can take. To analyze \eqref{aa2} let us replace the
variable $\mu$ by the variable
\be
u=\frac{1}{4 \pi} \int_{\mu}^{\mu_0} d\mu'/(\mu')^{1+\epsilon} \blue{,}
\ee
where $\mu_0$ is the bare IR cutoff, and we are interested in the limit $\mu \to 0$. In that
limit we see that $u \sim \mu^{-\epsilon}/\epsilon$ for $\epsilon>0$,
and $u= \frac{1}{4 \pi} \log(\mu_0/\mu)$ in $d=2$. As we pointed out in the main text,
the {running} derivative of the potential, which we denote $D_u(h)=V'(h)$ (indicating the explicit $u$ dependence)
satisfies (taking $\partial_h$ in \eqref{aa2}) a logarithmic PME
\be \label{RG2}
\partial_u D_u(h) = \partial_h^2 \log D_u(h)
\ee
As we will discuss in Sec. II this equation admits {the so-called self-similar solutions} of the form
\be \label{eqn:scaleW}
D_u(h) = u^{1- 2a} F( u^{-a} h),
\ee
where $F$ is a fixed function with $F(0)$ finite, and the exponent $a$ takes some given value (depending on $s$, see below).
{Since the self-similar solutions of the form \eqref{eqn:scaleW} lead to scale-invariant behavior and can act as attractors in the large $u$ limit, they can be referred to as fixed points of the RG flow.}

Recalling that $u \sim \mu^{-\epsilon}$ where $\mu \sim 1/L$ is the IR cutoff, we have $u \sim L^\epsilon$
where $L$ is the system size. 
{The form of the fixed point solution~\eqref{eqn:scaleW} implies the scaling behavior}
\be   \label{eq:roughness-1}
h \sim u^a \sim L^\alpha \quad   \text{with} \quad \alpha = a \epsilon
\ee
Let us recall
that the bare two point height correlator reads in Fourier, from the second derivative $[S^{(2)}(0)^{-1}]_{hh}$
in \eqref{dynaction} (or equivalently from \eqref{eqmo1})
\be
\langle h_{k,\omega} h_{k',\omega'} \rangle_0 = \frac{2 {\sigma_0}}{ \omega^2 + \nu_0^2 k^4} \times (2 \pi)^{d+1} \delta^{(d)} (k+k') \delta(\omega+\omega') . 
\ee
The exact (renormalized) correlator is similarly obtained from the second derivative matrix $[\Gamma^{(2)}(0)^{-1}]_{hh}$ and
takes the scaling form (setting ${ \sigma}={\sigma_0}=1$) for small $k,\omega$
\be \label{hhexact}
\langle h_{k,\omega} h_{k',\omega'} \rangle \simeq \frac{1}{\nu(k)^2 k^4} {\cal G}\left( \frac{\omega}{\nu(k) k^2  }\right ) \times (2 \pi)^{d+1} \delta^{(d)} (k+k') \delta(\omega+\omega'),
\quad  \quad {\cal G}(y)=\frac{2}{y^2 + 1} + O(\epsilon),
\ee
{where as follows from \eqref{eqn:scaleW}  }
\be
\nu(k) \sim \nu|_{\mu \sim k} =  D_u(0)|_{u \sim k^{-\epsilon}} \sim k^{- (1- 2 a) \epsilon}.
\ee
This implies that the dynamical exponent defined by $\omega \sim k^z$ is
\be
z = 2 - \epsilon(1- 2 a) = d + 2 \alpha
\ee
and obeys the general relation given in the text.
Furthermore from \eqref{hhexact} we see that in real space the equal time correlation
\be 
\langle h(x,t) h(x',t) \rangle = \int \frac{d \omega}{2 \pi} \int \frac{d^d k}{(2 \pi)^d} \frac{1}{\nu(k)^2 k^4} {\cal G}\left( \frac{\omega}{\nu(k) k^2  } \right )
\sim L^{2 a  \epsilon } \sim L^{2 \alpha}
\ee
consistent with the prediction for the roughness exponent given by \eqref{eq:roughness-1}.

{

\section{Functional RG flow: fixed points and their stability} 
\label{sec:RG-flow}



The RG flow equation~\eqref{RG2},
\begin{equation} \label{eq:Z}
\partial_u D_u(h) = \partial_h^2 \log D_u(h),
\end{equation} 
is itself a logarithmic fast diffusion equation, which represents a limiting case ($s \to 0$) of the fast diffusion equations, i.e. PMEs with $s < 1$,
see Chap 8. in \cite{vazquez2006smoothing} for details and applications. 
The solutions of these non-linear diffusion equations are expected to asymptotically approach, for large $u$, a self-similar solution,  independently of the initial condition  $D_0(h)$. In the RG picture this self-similar solution corresponds to a fixed point. We consider the forward or type I self-similar solutions of the form \eqref{eqn:scaleW},
\be
D_u(h)= u^{1-2 a} F_u(h u^{-a})
\ee
with $F_u(H)$ independent of $u$ for large $u$, which we denote $F(H)$.
The RG flow for $F_u(H)$ reads
\begin{equation} \label{eqn:FF2}
u \partial_u F_u(H) = a H F_u'(H) - (1-2a) F_u(H) + \frac{F''_u(H)}{F_u(H)} - \frac{F'_u(H)^2}{F_u(H)^2}
\end{equation}
where primes mean the derivative w.r.t. $H$. A self-similar solution is given by function $F(u)$  which satisfies Eq.~\eqref{eqn:F},
\begin{equation} \label{eqn:FF2-2}
a H F'(H) - (1-2a) F(H) + \frac{F''(H)}{F(H)} - \frac{F'(H)^2}{F(H)^2}=0.
\end{equation}
Note that if $F(H)$ is a solution of this equation, then $b^2 F(b h)$ is also a solution for any $b$. 

For the problem studied in this paper we will eventually restrict to solutions where $F(H)$ is finite and positive for all $H$ and even in $H$. 
However we will first study the most general solutions of \eqref{eqn:FF2-2}. 

\subsection{Classification of fixed points: Phase-plane formalism}
\label{sec:phase-plane-formalism}

In order to classify all possible self-similar solutions and identify their properties, we can apply the phase-plane formalism
see e.g. Section 3.8.2. in \cite{vazquez2006smoothing}. 
To that end, we rewrite equation~\eqref{eqn:FF2-2} for $F(H)$ as an autonomous system of two first-order differential equations for the functions $X(r) = \frac{H F'}{F}$ and  $Y(r)=H^2 F$,
which does not explicitly depend on $r = \ln H$,
\bea
\dot{X} &=& X + (1-2a-a X)Y, \label{eqs-XY0} \\
\dot{Y} &=& (2+X) Y, \label{eqs-XY2}
\eea
where $\dot{X} = \frac{d}{d r} X$ and so on. Each solution to the system~(\ref{eqs-XY0})-(\ref{eqs-XY2}) corresponding to a self-similar solution $F(H)$ can be represented by an integral curve in the phase plane $(X,Y)$.
Only one integral curve passes through each point in the phase plane. The system~(\ref{eqs-XY0})-(\ref{eqs-XY2}) has several singular points in  the phase plane $(X,Y)$ and each solution corresponds to an integral curve connecting two of them. Expansion around  singular  points  provides us the  asymptotic
behavior  of  $F(H)$  for  $H \to 0$ (in the case of starting point) and $H \to \infty$ (in the case of ending point).  Thus the analysis of the singular points of the system gives exhaustive  list  of  all  fixed points with their asymptotic behavior for large and small $H$.

One can rewrite the system~~(\ref{eqs-XY0})-(\ref{eqs-XY2}) as a single differential equation,
\bea
\frac{dY}{dX} = \frac{(2+X) Y}{X + (1-2a-a X)Y }. \label{eqs-XY-2}
\eea
We consider the fixed points with $F>0$, and thus, we look only at the integral curves in the upper plane $Y>0$. The integral curves connecting different singular
points in the phase plane $(X,Y)$  for $a=1/3$ and $a=-1/3$ are shown in Fig.~\ref{phase-plane-flow}.
The system has two finite singular points in the plane $(X,Y)$: $A=(0,0)$ and $B=(-2,2)$  whose position  does not depend on $a$, and three points
at infinity: $C=(-2+\frac{1}{a},\infty)$, $D=(-\infty,0)$, $E=(-\infty\ \mathrm{sign}(a),\infty)$.

\begin{figure}
{\includegraphics[width=7cm]{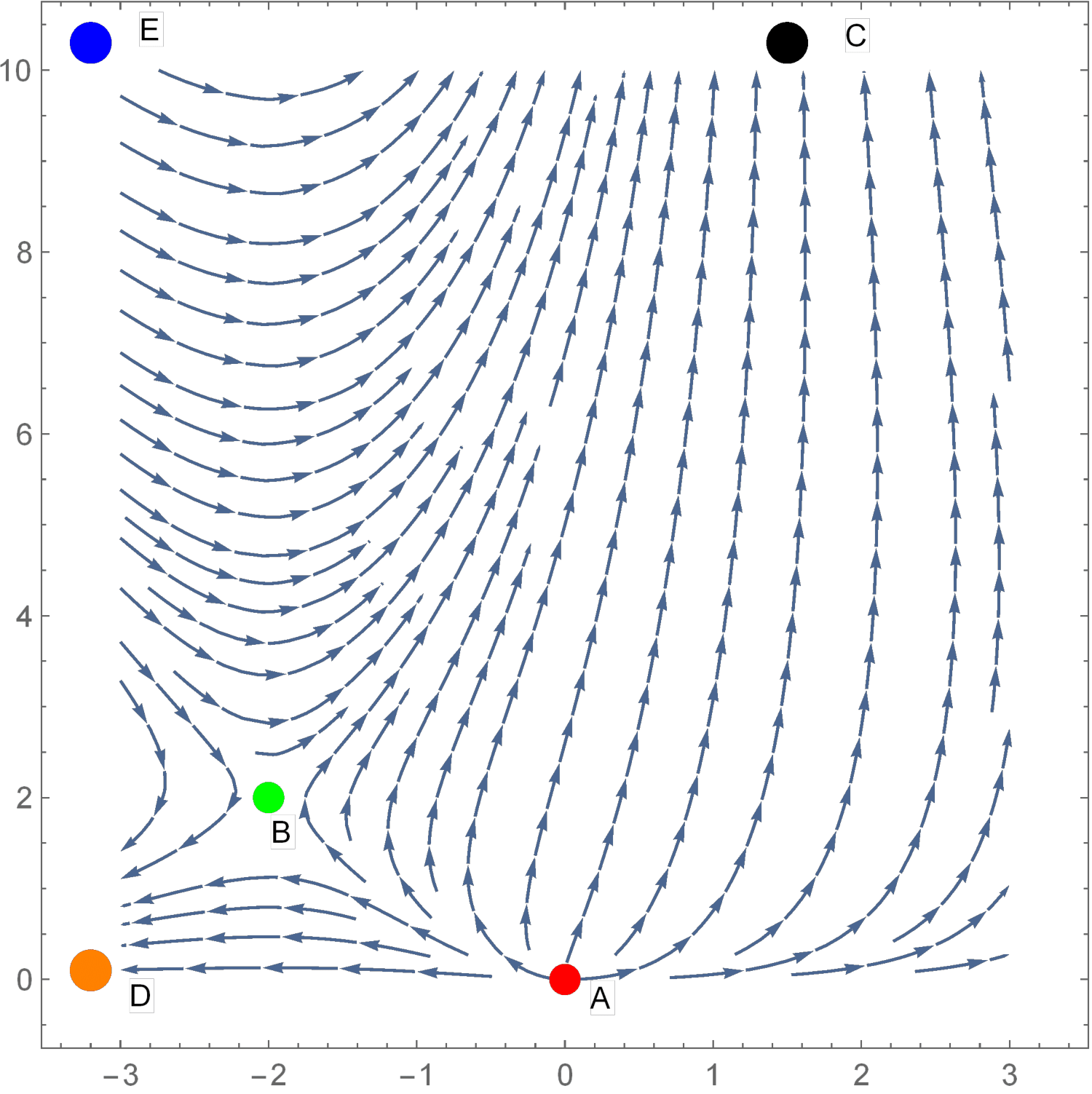} \ \
\includegraphics[width=7cm]{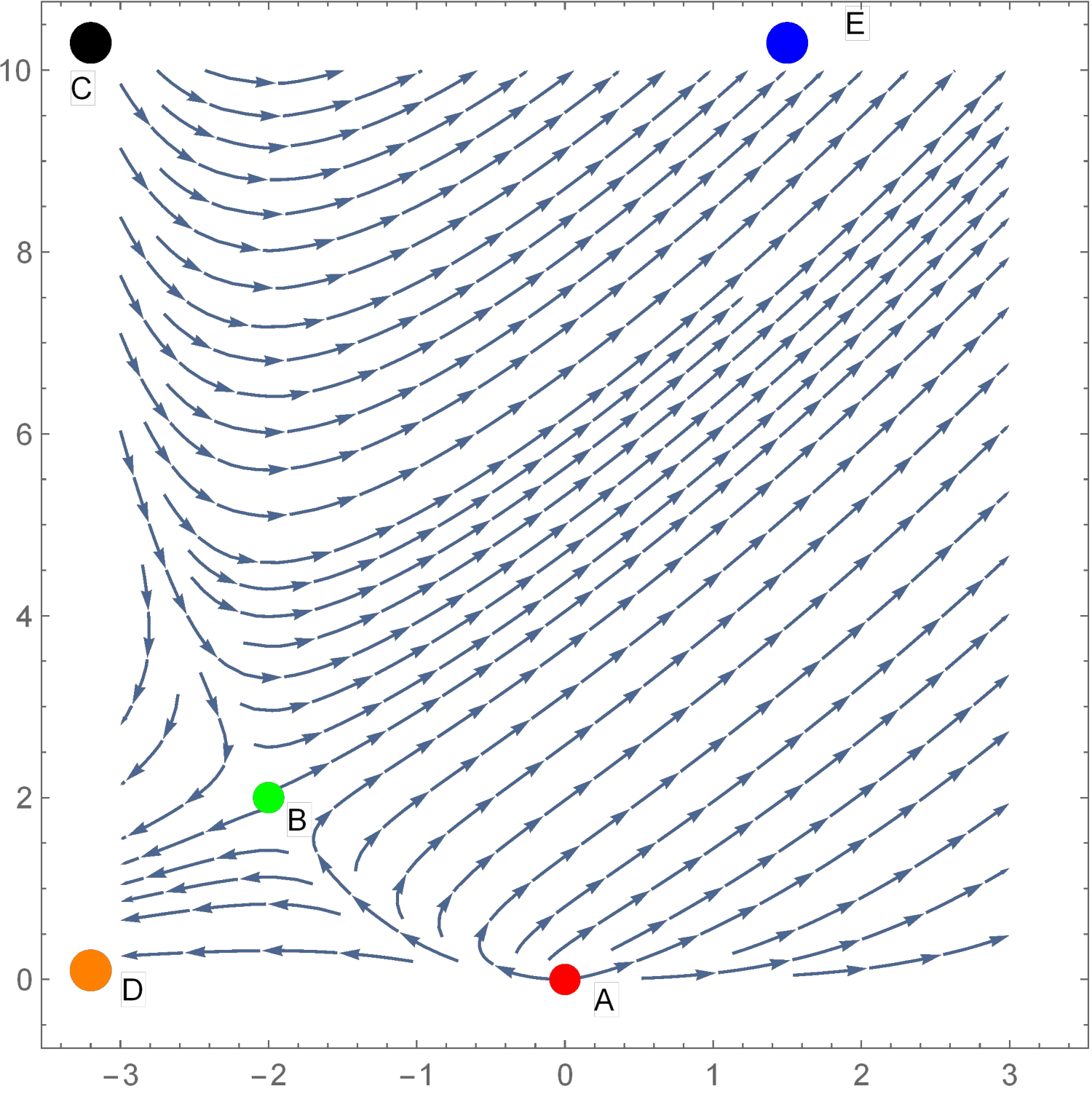}}
\caption{ 
Integral curves in the phase-plane  $(X,Y)$. Left panel for $a=1/3$ and right panel for $a=-1/3$.
There are 5 singular points A,B (in the plane) and C,D,E (at infinity). Note that for $a<0$ and $a>0$  the points C and E exchange their positions. Each curve connecting two singular points corresponds to a self-similar solution $F(H)$. The behavior at starting singular point gives asymptotic of $F(H)$ for $H \to 0$ and at the ending point for $H \to \infty$. Only curves starting at point A  give the solutions finite at $H=0$.   }
\label{phase-plane-flow}
\end{figure}

We now study the asymptotic behavior at each of these singular points, and identify among them repeller and sink points which can be only starting or ending points
for the integral curves.

(A) Expanding the system of equations~(\ref{eqs-XY0})-(\ref{eqs-XY2}) around point A we obtain
\bea
\dot{X} &=& X + (1-2a)Y, \label{eqs-XY-A-0} \\
\dot{Y} &=& 2 Y. \label{eqs-XY-A1}
\eea
Diagonalizing it, we find the eigenvalues $\lambda_{1,2}=\{1,2\}$.  Thus the point A is a repeller and the integral curves can only start
at it. Equation~(\ref{eqs-XY-A1}) implies  $Y(r) = C H^2$ for small $H$, where $C$ is a constant. This leads to  $F(H) = C$ at $H=0$
for all solutions which start at point A.

(B) Expanding around point B: $X=-2+x$ and $Y=2+y$ we obtain to linear order
\bea
\dot{x} &=&  (1-2a)x +y , \label{eqs-XY-B-0} \\
\dot{y} &=& 2 x. \label{eqs-XY-B}
\eea
Computing the eigenvalues we find 
$\lambda_{1,2}=\left\{\frac{1}{2} \left(-\sqrt{(2 a-1)^2 +8}-2 a+1\right)<0,\frac{1}{2} \left(\sqrt{(2 a-1)^2 +8}-2 a+1\right)>0\right\}$
for all $a$.
Thus the point B is a saddle point: there are integral curves starting and ending at the point B.
Close to B the solution behaves as $F(H)= \frac{Y}{H^2} = \frac{2}{H^2} + \frac{y(H)}{H^2}$ for small $H$ if it is a starting point and for large $H$
is it an ending point.

(C) Close to the singular point C the solution has the form
\bea
X  & \approx & X_c  = (1-2a)/a, \\
Y &=& C \exp ((2+X_c)r) =  C H^{2+X_c} . \label{eqs-XY-C}
\eea
This point is a sink point for $a>0$ with the asymptotic behavior $F(H) = C H^{X_c}$ for large $H$, and a repeller point  for $a<0$
with $F(H) = C H^{X_c}$ for small $H$. In the latter case the solution $F(H)$ diverges at the origin since $X_c<-2$ for $a<0$.

(D) Close to the point D the system can be simplified to
\bea
\dot{X} &=&  X , \label{eqs-XY-D-0} \\
\dot{Y} &=& X Y. \label{eqs-XY-D}
\eea
This point is a sink. The solution in its vicinity  is given by $X(r)=-C_1 e^{r}$ and $Y(r)=C_2 \exp[-C_1 e^{r}]$ with constants $C_{1,2}>0$.
The corresponding solutions have the asymptotic behavior of the form $F(H)\sim \frac{ e^{-C_1 H}}{H^2}$ for large $H$.

(E)  Close to the point E the system can be simplified to
\bea
\dot{X} &=&  -a X Y , \label{eqs-XY-E-0} \\
\dot{Y} &=& X Y, \label{eqs-XY-E1}
\eea
whose solution, $X(r) =  C_1 /(1-e^{C_1 r-C_2 }) $ and 
$Y(r)=C_1/[a(1-e^{-C_1 r+C_2 })]$, diverges at finite $r_0=C_2/C_1$.
The singular point E is a sink  for $a<0$ and a repeller for $a>0$. 
Thus, for $a<0$ the corresponding self-similar solution  $F(H)=Y(\ln 
H)/H^2$ is defined  for $H \in [0,H_0)$ with $H_0=e^{r_0}$ and diverges 
in the limit $H \to H_0^-$, while for $a>0$ it is defined  for $H \in 
(H_0,\infty)$ and diverges in the limit $H\to H_0^+$.
The above form of $X(r)$ and $Y(r)$ implies the asymptotic behavior 
$F(H) \simeq -\frac{C_1}{a H^2 
\left(\left(\frac{H_0}{H}\right)^{C_1}-1\right)}$ slightly below 
$H_0=e^{C_2/C_1}$ for $a<0$ and above $H_0$ for $a>0$. Expanding it in a 
Laurent series near the pole $H=H_0$, we obtain
$F(H) = 1/[a H_0 (H-H_0)] + O\left((H-H_0)^0\right)$.
The coefficient in the leading term is exact, as we checked numerically
at fixed $a<0$. 

For $a=0$ the system (\ref{eqs-XY0})-(\ref{eqs-XY2}) simplifies to
\bea
\dot{X} &=&  X + Y , \label{eqs-XY-E-a0} \\
\dot{Y} &=& (2+X) Y, \label{eqs-XY-E1-a0}
\eea
whose solution is  $X(r)=\pm e^{r} \sqrt{C_1} \tan \left[\frac{1}{2} \sqrt{c_1} \left(C_2\pm e^{r} \right)\right]$ and
$Y(r)=\frac{1}{2} e^{2r} C_1 \sec^2\left[\frac{1}{2} \sqrt{C_1}
   \left(C_2 \pm e^{r} \right)\right]$. This leads to an exact self-similar solution $F(H)=\frac{1}{2} C_1 \sec ^2\left[\frac{1}{2} \sqrt{C_1}(C_2\pm H)\right]$. This solution is negative for $C_1<0$ and decays as $F(H) \sim - \exp[-\sqrt{|C_1|}H]$ for large $H$. For $C_1>0$ it diverges at finite $H_0=\pm \frac{\sqrt{C_1} C_2+  (2n-1)\pi  }{\sqrt{C_1}}$, $n \in \mathbb{Z}$ as $F(H) = 2/{(H-H_0)^2} + O\left((H-H_0)^0\right) $.
\\

We are now in the position to analyze all possible self-similar solutions. Here we restrict ourselves to the solutions which are finite in the vicinity of the origin $H=0$. Only the solutions represented by the integral curves starting at point $A$ satisfy this condition. There are four different classes of such solutions
corresponding to the integral curves connecting the point A with points B,C, D or E:

($A \to B$) This solution decays  as $F(H) \sim 1/H^2$ for large $H$.

($A \to C$) This solution exists only for $a>0$ and behaves as  $F(H) \sim H^{-2 + 1/a}$ for large $H$.

($A \to D $) This solution decays as $F(H)\sim \frac{ e^{-C_1 H}}{H^2}$ for large $H$.

($A \to E $) This solution exists only for $a \leq 0$ and diverges at finite  $H=H_0$.
}
\\

Let us give here some additional properties, as well as a few exact solutions 
which can be obtained for some values of $a$ 

For $a=1$ there is a special solution
\be  \label{particular} 
F(H) = \frac{1}{1 + \frac{1}{2} H^2} 
\ee 
which is a limiting case of a more general family of solutions for $a=1$ 
\be 
F(H)= \frac{c_1^2}{c_1^2 c_2 e^{c_1 H}-c_1 H-1}
\ee 
which however are not even in $H$, except in the limit $c_1 \to 0,c_2 =1/c_1^2 + 1$
where one recovers \eqref{particular}. 

For $a=0$ there is a solution of the form
\be 
F(H)=\frac{1}{2} \left(1+\tan\left(\frac{H}{2}\right)^2\right) 
\ee 
which is even but diverges at $H= \pi$. As discussed above, there are additional solutions, which however
are not even or not positive.

{\bf Remark}. For $0<a<1$ the fixed point $F(H)$ behaves at large $|H|$ as $F(H) \sim |H|^{s-1}$ with $a=1/(1+s)$. One can check that
it takes the form $F(H)=|H|^{s-1} g(1/|H|^{s+1})$, where the function $g(z)$ satisfies a closed differential equation.
Solving in an expansion at large $|H|$, one finds $g(z)=1 + \sum_{n \geq 1} g_n z^n$ with $g_1=1-s$, $g_2= \frac{1}{2} (1 - s) (1 + s) (2 + s)$,
and so on (up to the rescaling $b^2 F(b H)$). Most notably we find the leading correction is always $(s-1)/H^2$, with a
universal coefficient (i.e independent of $b$). This remark helps to understand the structure of the flow towards the fixed point,
see below.



\subsection{Numerical calculation of fixed points}

We now study numerically the solutions of \eqref{eqn:FF2-2}. Since the flow \eqref{eq:Z} preserves the parity of $D_u(h)$, and the initial conditions $D(h) = D_0(h)$ considered in this paper are even functions, it follows that 
we should focus on $F(H)$ even, positive, with $F(0)$ finite 
and $F'(0) = 0$.
Moreover, if $F(H)$ is a solution to Eq.~\eqref{eqn:FF2-2}, then $b^2 F(H b)$ is also a solution. Hence, we can impose the condition $F(0) = 1$.
To perform numerical integration, we first transform equation \eqref{eqn:FF2-2} into a system of two first-order differential equations for $F$ and $G:=F'$:

\begin{align}
&G'(H) = \frac{G(H)^2}{F(H)} + (1-2a)F(H)^2 - aH G(H) F(H)\\
&F'(H) = G(H).
\end{align}
Starting from the chosen initial conditions, $F(0)=1$ and $G(0)=0$, we employ a fifth-order Runge-Kutta integration scheme to solve the system numerically.
 
According to the classification of possible fixed points performed  in Sec.~\ref{sec:phase-plane-formalism} we expect four different types of solutions. 
In Fig. \ref{fig:figFH} Left, we recover two of them: 
\begin{itemize}
    \item the class ($A \to E $). For $a<0$, $F(H)$ diverges at finite $H$, and for $a=0$,  we recover the exact solution $F(H)=(1+\tan(H/2)^2)/2$.
    \item the class ($A \to C$). Shown here for $0 < a < 1/2$, the solutions exhibit an asymptotic behavior $F(H) \sim |H|^{-2+1/a}$.
\end{itemize}

In Fig. \ref{fig:figFH} Right, we recover the remaining cases:
\begin{itemize}
    \item the class ($A \to C$). Shown here for $1/2<a<1$ when $F(H) \sim |H|^{-2+1/a}$ decreases with $H$
    \item the class ($A \to B$). For $a=1$, we recover the exact solution $F(H) = 1/(1+H^2/2) \sim 1/H^2$.
    \item the class ($A \to D$). For $a>1$, we find solutions that decay exponentially.
\end{itemize}

Note that the behavior of $F(H)$ changes qualitatively at $a=0$ and $a=1$.



\begin{figure}
    \centering
    \includegraphics[width=0.48\textwidth]{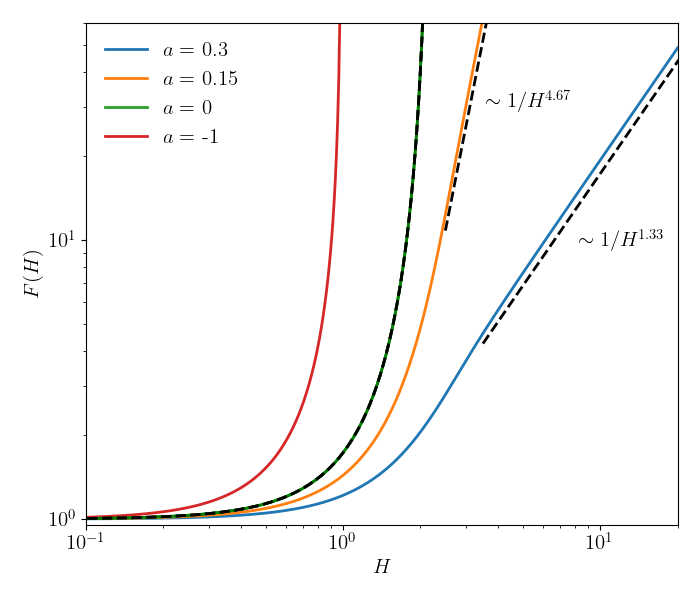}
    \includegraphics[width=0.48\textwidth]{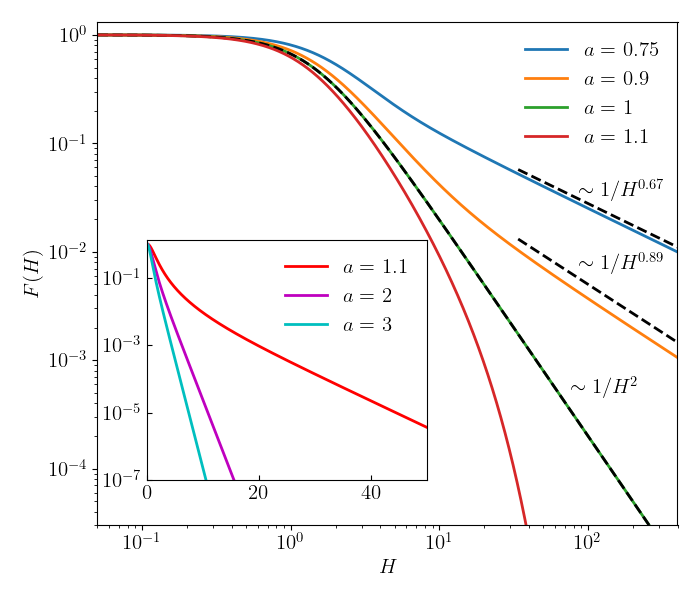}
    \caption{ Solution $F(H)$ of \eqref{eqn:FF2} for different values of $a$. Left: $a<1/2$. For $0<a<1/2$, the solution shows an asymptotic behavior $\sim H^{-2+1/a}$
    at large $H$, while it diverges at finite $H=H_0$ as $\sim 1/(H_0-H)$ 
for $a < 0$ and $\sim 1/(H_0-H)^2$ for $a=0$. Right: $a>1/2$, For $1>a>1/2$, $F(H)$ decreases at large $H$ as $\sim H^{-2+1/a}$. At $a=1$, the asymptotic behavior changes and $F(H) = 1/(1+H^2/2) \sim 1/H^2$. For $a>1$, we show in the inset that the semilogarithmic plot of $F(H)$ is well fitted by a straight line at large $H$, thus $F(H)$ decays exponentially in $H$. 
}
    \label{fig:figFH}
\end{figure}

\subsection{Linear stability analysis} 


Finding the fixed points alone is insufficient, as they may not be relevant. Therefore, we analyze their stability to determine whether the fixed points are attractive.
Here will restrict the analysis to the power law case $0<a<1$ ($A \to C$) , although
some of the equations below are more general. We examine the linear stability of the fixed point $F(H)$ using the flow equation \eqref{eq:Z}.
To achieve this, we linearize the equation \eqref{eq:Z} around the fixed point $F(H)$ by introducing a perturbation $\Phi$, defined as:
\begin{equation} \label{eq:pertF}
D_u(h) = u^{1-2a}F_u(H=u^{-a} h) = u^{1-2a} (F(H) + \delta F_u(H) ) \quad , \quad  \delta F_u(H) = \Phi (H) e^{\lambda \ln{u}},
\end{equation}
where $\lambda$ is the eigenvalue that we aim to estimate.
We must restrict to perturbations $\Phi(H)$ which are smaller or at most of order $F(H) \sim |H|^{ \frac{1}{a}-2}$ at large $|H|$.
Indeed, if it is not the case these will flow to another fixed point characterized by a smaller value of $a$.
The fixed point will be stable if no such perturbation exist with a strictly positive eigenvalue.
By substituting \eqref{eq:pertF} into the flow equation \eqref{eq:Z} and expanding to linear order in $\Phi$, we derive the following linear differential equation for $\Phi$:
\begin{equation} \label{eqd2H2}
0 =  ( {\cal L}_\lambda  \Phi)(H)  =  \Phi''(H) - \Phi'(H) \left(2 \frac{F'(H)}{F(H)}-a H F(H)\right) + \Phi(H)\left[a H F'(H) -(2 - 4a + \lambda)F(H) + \frac{F'(H)^2}{F(H)^2}\right].
\end{equation}
Equivalently, the flow of a perturbation of the fixed point $F_u = F + \delta F_u$ is given by $u \partial_u \delta F_u = \frac{1}{F} {\cal L}_0 \delta F_u$
and we want to find the spectrum of ${\cal L}_0$, i.e. its eigenvalues $\lambda$ and associated eigenvectors $\Phi$,
such that $\delta F_u = u^\lambda \Phi$.

We have studied the equation \eqref{eqd2H2} analytically and numerically. From parity one must choose $\Phi'(0)=0$, and since it is a linear equation we can choose $\Phi(0)=1$.
For the numerical integration, we simultaneously solve for $F$ and $\Phi$ by converting the two second-order equations into a system of four first-order equations for $F$, $F'$, $\Phi$, and $\Phi'$. As previously done for $F$, we use a Runge-Kutta integration scheme.
The only parameter is $\lambda$.

We have found the following results:

\begin{itemize}

    \item There are two exact eigenvectors related to symmetries: The first symmetry is  $F \to b^2 F(b h)$, which gives the exact eigenvalue $\lambda=0$  corresponding to the eigenvector  $\Phi(H) = F(H)+\frac12 H F'(H)$ (for an analogous analysis see e.g. Eqs (S.29 ) and (S.32) in \cite{balogdisorderarxiv,balog2018disorder}).
    It behaves at large $|H|$ as $\Phi(H) \sim |H|^{\frac{1}{a} - 2}$.

    \item The second symmetry is  $u \to u+u_0$, which gives the exact  eigenvalue $\lambda=-1$   corresponding to the eigenvector  $\Phi(H) = F(H)-\frac{1}{1/a-2} H F'(H)$. It behaves at large $|H|$ as $\Phi(H) \sim 1/H^2$.

\item There is a continuum spectrum with eigenvectors which behave as power laws $\Phi(H) \sim |H|^{\frac{1}{a} - 2 - \delta}$
with eigenvalue $\lambda = - a \delta$ and $\delta>0$. This value of $\lambda$ can be simply verified by injecting the
large $|H|$ behavior in \eqref{eqd2H2}. The existence of these eigenvectors is easily checked numerically.

\item In addition to the continuum spectrum there is a discrete spectrum which correspond to eigenvectors with fast decay at large $|H|$
(stretched exponential). The most relevant one has eigenvalue $\lambda = a- 1$. It can be found exactly as follows. Consider the
change of function
\be
\Phi(H) = F(H) e^{- \frac{1}{2} \int^H a H F }  \Psi(H)
\ee
Then \eqref{eqd2H2} is equivalent to the Schrodinger problem
\bea
&& 0 = {\cal H}_\lambda \Psi = - \Psi'' + V_\lambda(H) \Psi \\
&&
V_\lambda(H) = W'(H) + W(H)^2   + (\lambda - (a-1) ) F  \quad , \quad W(H) = -  \frac{a}{2} H F(H)
\eea
Hence $\Psi(H)$ must be a zero energy eigenvector of the Schrodinger problem in the potential $V_\lambda(H)$.
For the case $\lambda=a-1$, we see that $V_\lambda(H)$ has the familiar form which arises in supersymmetric quantum mechanics,
and an eigenvector is then
\be
\Psi(H) = e^{ - \frac{a}{2} \int^H dH' H' F(H')}  \quad , \quad \Phi(H) =  F(H) e^{ - a \int^H dH' H' F(H') }
\label{eq:eigenvecta1}
\ee
which has a fast decay at infinity. Hence we have found an exact eigenfunction $\Phi(H)$, associated to
the eigenvalue $\lambda=a-1$, of our original problem. Interestingly it behaves at large $|H|$
as $|H|^{s-1} \exp( - \gamma |H|^{s+1} ) $, reminiscent of the behavior of $P(h)$ in \eqref{PDFhcont}
(although we could not see any obvious relation).  Next, since one can write
${\cal H}_\lambda  = Q^\dagger Q   + (\lambda - (a-1) ) F  $
with $Q=\partial_H - W $, we see that no fast decaying eigenvector can exist for $\lambda > a-1$
since then ${\cal H}_\lambda$ must have a positive spectrum.
Finally note that in the $\Psi$ variable the linearized flow
equation reads
$u \partial_u \Psi = - \frac{1}{F} {\cal H}_0 \Psi$.

We have confirmed the above result by computing numerically the eigenvector for $\lambda=a-1$, and we have other
fast decaying eigenvector $\Phi$ for a set of discrete values of $\lambda$, which are all smaller than $a-1$.
In order to identify these eigenvalues numerically, we note that $\Phi(H)$ is a smooth function of $\lambda$ as the differential equation is linear. Therefore, when $H \to +\infty$, we have $\Phi(H) \sim \frac{A(\lambda)}{H^w}$,
where $A(\lambda)$ is a smooth function of $\lambda$.
Therefore, to find a function \(\Phi\) with exponential decay, $A(\lambda)=0$ must hold. One can then assume that $A$ changes sign near this root.
\end{itemize}
{Finally, we have shown that, for $0<a<1$, eigenvectors are associated to negative eigenvalues, provided they decay faster than $F(H)$ at large $H$. Hence these fixed points are stable.}


\begin{figure}
    \centering
    \includegraphics[width=0.48\linewidth]{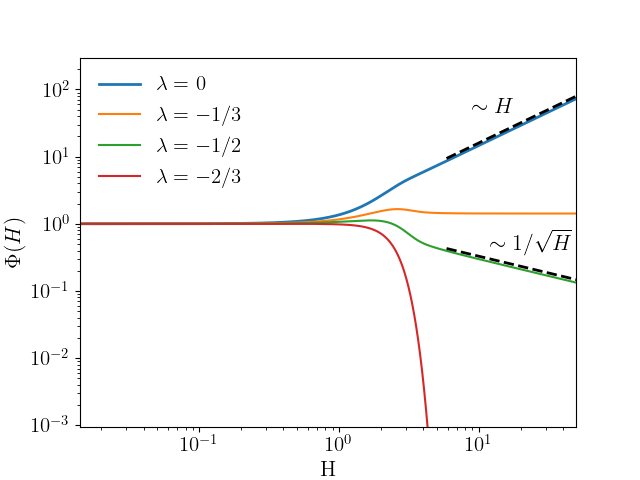}
    \includegraphics[width=0.48\linewidth]{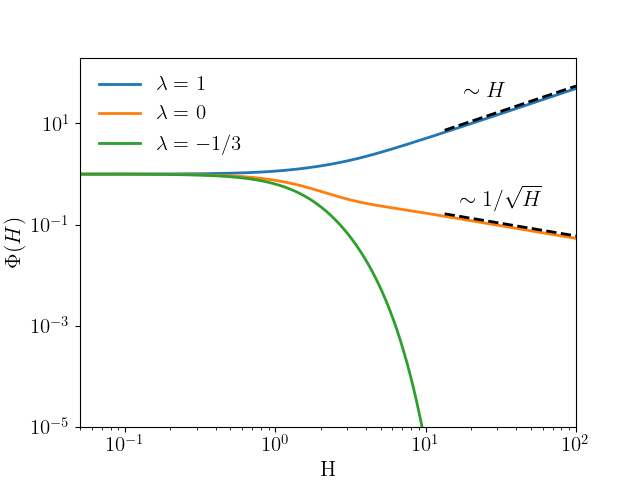}
    \caption{(Left) Plot of $\Phi(H)$ for $a=1/3$, $s=2$ , and several $\lambda$. For $\lambda= 0, -1/3,-1/2$, one retrieves the power law eigenvectors with $\Phi(H) \sim |H|^{-2+1/a + \lambda/a}$ for large $H$. For $\lambda=a-1=-2/3$, we retrieve an eigenvector with fast decay as predicted. (Right) Same for $a=2/3$, $s=1/2<1$. We also retrieves the power law eigenvectors, and the fast decay occurs here at $\lambda=a-1=-1/3$}
    \label{fig:phi}
\end{figure}

\subsection{Basin of attraction}

We previously showed that the fixed points found of the form  $D_u(h) = u^{1- 2 a} F(u^{-a} h)$ are stable {\green for $0<a<1$}. A more precise test consists in starting from the initial conditions $D(h)=D_0(h)$ and checking that the RG flow indeed converges towards the self-similar solutions obtained earlier. To evaluate numerically the flow equation

\begin{equation}
\partial_u D_u(h) = \partial_h^2 \log D_u(h),
\label{eq:eqflowSM}
\end{equation}

we consider a large finite grid $\{h_i\}$ of spacing $dh$: $\{-L_R,-L_R+dh, \dots, L_R-dh,L_R\}$. We evaluate the discrete Laplacian using $(f(h_{i+1})+f(h_{i-1})-2 f(h_{i}))/dh^2$. At the boundaries, we take the Laplacian to be equal to the one evaluated in respectively $2$ and $N-1$. We then integrate with respect to $u$ using a Euler scheme with an integration step $du$.

We focus on the case $s>0$. In particular, in the main text, we considered $s=1/2$ and $s=2$. For $s=2$, we take the initial condition $D_0(h) = 2 \sqrt{4 +h^2}$, which is
analytic in $h=0$,
$L_R=50$ and $du=3e-5$. For $s=1/2$, we consider $D_0(h) = 1/2(1/4^4+h^2)^{-1/4}$ and $L_R=50$. For large $h$, since one has $D_0(h) \sim 1/\sqrt{|h|} \ll |\log{(D_0(h))}| \sim |\log{|h|}|$, it is necessary to use a smaller integration step $du=5e-7$.
In Fig. \ref{fig:figRG} of the main text, we show that, in both cases, under the RG flow equation \eqref{eq:eqflowSM} the asymptotic behavior at large $|h|$ is preserved,
and that $F_u(H)$ converges towards the self-similar fixed point $F(H)$ with the asymptotic behavior ($A \to C$), and the corresponding parameter $a=\frac{1}{1+s}$.
This confirms that the fixed points studied previously are stable, and more importantly, that they attract the functions $D(h)=D_0(h)$ of the form chosen in this paper.

Thanks to linear stability analysis, we can elucidate the approach to the fixed point, $F_u(H)-F(H) \sim u^\lambda \Phi(H)$ by determining the largest eigenvalue 
$\lambda$ and the form of its corresponding eigenvector $\Phi(H)$. Apart from a discrete set of eigenvectors that decay exponentially fast, with the largest eigenvalue being $\lambda_{\rm exp}=a-1$, all other eigenvectors exhibit power-law decays. There are potentially two relevant power-law eigenvectors.
\begin{itemize}
    \item Starting from initial conditions of the form $D_0(h) \simeq C_1 h^{s-1} + C_2 h^{s-1-\gamma}$, the second term will be corrected during the flow and decay with eigenvalue $\lambda_\gamma = -a \gamma = -\gamma/(s+1)$ (given that $0 < a < 1$).
    \item Recall that the fixed point takes the form $F(H) \sim H^{s-1} + g_1/H^2$. The second power law will arise during the flow and is associated to the eigenvalue $\lambda_F = -1$.
\end{itemize}
Therefore, the largest eigenvalue is either $\lambda_{\rm exp}$, $\lambda_\gamma$ or $\lambda_F$. However, the latter is always irrelevant since $\lambda_F = -1 < a - 1 = \lambda_{\rm exp}$. Ultimately, we find that the largest eigenvalue and the form of its corresponding eigenvector depend on the values of $\gamma$ and $a = 1/(1+s)$. For $\gamma < s$, $\lambda_\gamma = -a \gamma > a - 1 = \lambda_{\rm exp}$, and the largest eigenvalue is $\lambda_\gamma = -\gamma/(s+1)$, leading to $\Phi(H) \sim H^{s-1-\gamma}$ decaying as a power-law. For $\gamma > s$, the largest eigenvalue is $\lambda_{\rm exp} = a - 1$, and the leading eigenvector $\Phi(H)$ decays exponentially fast in $H$. This is illustrated in Fig.~\ref{fig:flowRGdiff}.
On the left, starting from initial conditions $D_0(h) = \frac{1}{2 \left(1/4^4 + h^2 \right)^{1/4}} \simeq 2 h^{-1/2} - h^{-5/2}/2^9$, corresponding to $s = 1/2$ and $\gamma = 2$, we expect the largest correction of $|F(H) - F_u(H)|$ to scale as $u^{a-1} = u^{1/3}$. The eigenvector $\Phi(H)$ is given by \eqref{eq:eigenvecta1}, up to a constant, chosen to match the behavior at $H = 0$. $\Phi(H)$ indeed decays exponentially in $H$, although there is a slight difference with the analytical solution, likely caused by the finite $u$.
On the right panel, the initial condition is $D_0(h) = 1 + h^2$, corresponding to $s = 3$ and $\gamma = 2$. In this case, $\gamma < s$, and the largest eigenvalue is $-a \gamma = -1/2$. Its corresponding eigenvector behaves as $\Phi(H) \sim 1$ in the limit $H \to \infty$. Both these properties are checked numerically in Fig.~\ref{fig:flowRGdiff}.



\begin{figure}
    \centering
    \includegraphics[width=0.48\linewidth]{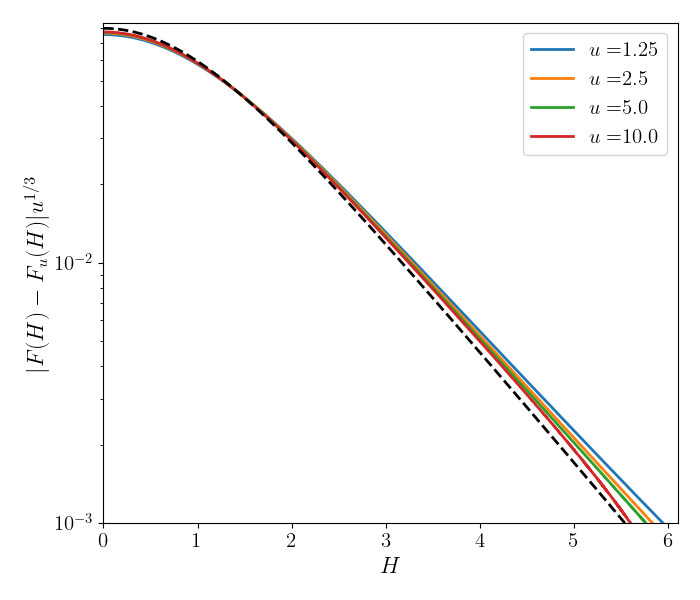}
    \includegraphics[width=0.48\linewidth]{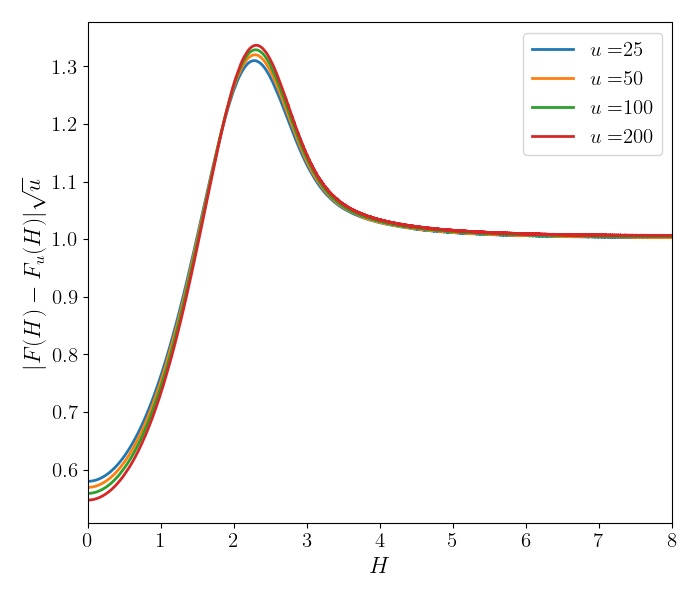}
    \caption{RG flow near the fixed point $|F(H)-F_u(H)|$ rescaled by $u^{\lambda}$, with $\lambda$ the predicted largest eigenvalue. 
    (Left panel) Starting from $D_0(h) = \left(1/4^4 + h^2 \right)^{-0.25}/2$ ($s=0.5$), with $\lambda=a-1=-1/3$.  Its corresponding eigenvector in black dotted lines decays exponentially (Right panel) Starting from $D_0(h) = 1+h^2$ ($s=3$). One has $\lambda=-1/2$ with its corresponding eigenvectors of order $1$ when $H \to \infty$ }
    \label{fig:flowRGdiff}
\end{figure}

\section{Predictions from the random walk model} \label{secPredRW}

The random walk model allows for very detailed predictions. We will study the discrete random walk model
but at large scale one can use a continuum description which we will also study. 

\subsection{Second moment, scaling and anomalous scaling}

We consider the random walk model 
\be
h_{n+1}=h_n + \frac{1}{\sqrt{D(h_n)}} \xi_n  
\label{RWmodelSM}
\ee
for $n \geq 0$, where the $\xi_n$ are i.i.d Gaussian random variables with unit variance. 
The trajectory of the walker can be identified with an interface in the stationary regime, and its position corresponds to the height of the interface.
We recall that at large $h$ one has $D(h) \simeq c |h|^{s-1}$. The increments being independent
the variance of the height differences is equal to
\be  \label{variancedifference}
\langle (h_{n+\ell} - h_n)^2 \rangle = \sum_{m=n}^{n+\ell-1} \langle \frac{1}{D(h_m)} \rangle 
\ee 
This relation is exact for any $n \ge 0$. 
We now consider a random walk with an initial position $h_0 = O(1)$.  At large $n$, we assume the scaling relation $h_n \sim n^\alpha \tilde{h}$, where $\alpha$ is the Hurst exponent and $\tilde{h}$ is the rescaled height, a random variable that is independent of $n$. The one point distribution of $\tilde{h}$ is discussed in Section \ref{secincr}. Now we determine the exponent $\alpha$ using a self-consistent argument.
Using \eqref{variancedifference} one obtains
\be 
\langle h_n^2 \rangle = \sum_{m=0}^{n-1} \langle \frac{1}{D(h_m)} \rangle
\ee 
For $s>0$ and for large $n$, we can estimate the r.h.s as 
\be 
 \sum_{m=0}^{n-1} \langle \frac{1}{D(h_m)} \rangle  \simeq \frac{1}{c} \sum_{m=0}^{n-1} \langle |h_m|^{1-s} \rangle
 \simeq \frac{1}{c} \sum_{m=0}^{n-1} m^{\alpha(1-s)} \langle |\tilde h|^{1-s} \rangle \simeq \frac{1}{c (1 + \alpha(1-s))} n^{1 + \alpha(1-s)}
 \langle |\tilde h|^{1-s} \rangle 
\ee 
while the l.h.s. behaves as $\langle h_n^2 \rangle \simeq \langle \tilde h^2 \rangle n^{2 \alpha}$.
Equating both determines $\alpha$ and gives a relation between the moments of $\tilde h$
\be \label{momentrelationRW}
\alpha = \frac{1}{1+s} \quad , \quad \langle \tilde h^2 \rangle = \frac{s+1}{2 c} \langle |\tilde h|^{1-s} \rangle 
\ee 
Note that the exponent of the random walk model coincides with our result for the global roughness exponent for the interface, given in the text. Furthermore, we now demonstrate that the random walk exhibits anomalous scaling behavior, as observed numerically for the stationary interface in Figure 3 of the main text. Let us evaluate the variance of the height difference using \eqref{variancedifference}.
In the right hand side we can write for $n \gg 1$ and any $\ell \geq 1$
\be  \label{sumscalemom2}
 \langle (h_{n+\ell} - h_n)^2 \rangle = \sum_{m=n}^{n+\ell-1} \langle \frac{1}{D(h_m)} \rangle  \simeq \frac{1}{c} \langle \tilde h^{1-s}\rangle \sum_{p=0}^{\ell-1} (n+p)^{(1-s)/(1+s)}  = \frac{1}{c} \langle \tilde h^{1-s}\rangle  \left(  \zeta
   \left(\frac{1-s}{s+1},n\right)-\zeta
   \left(\frac{1-s}{s+1},\ell+n\right) \right) 
\ee  
where $\zeta(a,x)$ is the Hurwitz zeta function. There are two regimes. 


In the first regime $\ell \ll n$, which is relevant for the study of local properties, one can 
neglect the dependence in $p$ in the sum in \eqref{sumscalemom2} and one finds an anomalous scaling:
\begin{equation} \label{eq:scalemom2}
\langle (h_{n+\ell} - h_n)^2 \rangle \simeq \frac{1}{c} \langle \tilde{h}^{1-s}\rangle n^{2 \alpha - 1} \, \ell
\end{equation}

This corresponds, for all $s > 0$, to a local roughness exponent for the random walk:

\begin{equation} \label{q2anomdisc}
\alpha_{\rm loc}(q=2) = \frac{1}{2}
\end{equation}

In the second regime, $\ell = O(n)$, one recovers the global scaling, more precisely
\be 
\langle (h_{n+\ell} - h_n)^2 \rangle \simeq \frac{1}{c} \langle \tilde h^{1-s}\rangle n^{2 \alpha } f(\lambda= \ell/n) \quad , \quad
f(\lambda)= \frac{s+1}{2} ( (1 + \lambda)^{2/(s+1)} - 1) 
\ee 
We observe that the two regimes match when $\lambda \to 0$.

\subsection{Distribution of the increments and multiscaling} \label{secincr}

In order to obtain the local roughness exponents $\alpha_{\rm loc}(q)$ for any $q$ and the multiscaling properties of the random walk, 
one must study the distribution of the increments. To this aim we more information about the distribution of the
scaled height $\tilde h= h_n/n^\alpha$. At large $n$ it becomes well described by a continuum
model for a height field $h(x)$ which obeys the Ito stochastic equation
$dh(x) = \frac{1}{\sqrt{D(h(x))}}  dB(x)$ where $B(x)$ is a Brownian motion. The
one point PDF of $h(x)$, $P(h,x)$ satisfies the Fokker-Planck equation
\be \label{FKPh} 
\partial_x P(h,x) = \frac{1}{2} \partial_h^2 (P(h,x)/D(h)) 
\ee 
This equation and the stochastic process are studied in Section, where we show that at large $x$, using $D(h) \simeq c |h|^{s-1}$ 
with initial condition $h_0=h(0)=O(1)$,
the PDF takes the scaling form
\bea \label{PDFhcont}
&& P(h,x) \simeq \frac{1}{2} \frac{1}{x^{1/(1+s)}} \tilde  P( |h|/x^{1/(1+s)}) \\
&& \tilde  P(\tilde h) =  
\mathcal{N} |\tilde h|^{s-1} e^{-\frac{2 c |\tilde h|^{s+1}}{ (s+1)^2}} 
\quad , \quad \mathcal{N}  = 
\frac{(2 c)^{\frac{s}{s+1}} s
   (s+1)^{-\frac{2 s}{s+1}}}{\Gamma
   \left(\frac{2
   s+1}{s+1}\right)}
\eea 
which is normalized as $\int_{-\infty}^{+\infty} dh P(h,x)=1$ while 
$\int_0^{+\infty} d\tilde h \tilde P(\tilde h) = 1$. One can check
explicitly that the moment relation in \eqref{momentrelationRW} holds exactly for
this scaled variable $\tilde h$. Hence we can identify this variable $\tilde h$
in the discrete and the continuum model.

\subsubsection{Distribution of the local increment } 

We start by computing the distribution of the increment of length $1$ in the discrete model 
$\Delta= h_{n+1}-h_n$ for large $n$. In this case, the discrete setting is particularly convenient as this increment is exactly
Gaussian for a given $h_n$ , hence its distribution is given by 
\be 
Q_n(\Delta) \simeq  \int_{-\infty}^\infty dh \sqrt{\frac{D(h)}{2\pi}}  e^{- \frac{ D(h) \Delta^2}{2} } P(h,n) \\  
\ee 
where we use the continuum formula \eqref{PDFhcont} for the PDF of $h_n$ with $x \simeq n$. Depending of 
the value of $\Delta$ the region in $h$ which dominate the integral are different. 
Let us consider first the case where the integral is dominated by values of $|h| \gg 1$
so that we can replace $D(h) \simeq c |h|^{s-1}$. Inserting 
\eqref{PDFhcont} we obtain
\bea  \label{eq:exprQdis}
&& Q_n(\Delta) \simeq  \mathcal{N} \sqrt{\frac{c}{2\pi}} \frac{1}{n^{s/(1+s)}} \int_0^\infty dh h^{\frac{3}{2} (s-1)} e^{- \frac{ c h^{s-1}  \Delta^2}{2} 
-\frac{2 c h^{s+1}}{n (s+1)^2}} 
\eea 
Upon rescaling $h = n^{1/(1+s)} \tilde h$ one obtains 
\bea \label{eq:eqforQtilde}
&& Q_n(\Delta) \simeq  n^{\frac{1}{2}-\alpha} \tilde Q( \Delta n^{\frac{1}{2}-\alpha} ) \\
&& \tilde Q( \tilde \Delta) =  \mathcal{N} \sqrt{\frac{c}{2\pi}}  
\int_0^\infty d\tilde h \tilde h^{\frac{3}{2} (s-1)} e^{- \frac{ c  \tilde h^{s-1} \tilde \Delta^2}{2} 
-\frac{2 c \tilde h^{s+1}}{ (s+1)^2}} \label{eq:expQint}
\eea  
For $s>1$ and in the large $\tilde \Delta$ limit, the integral is dominated by small $\tilde h$
and 
one can neglect the second term in the exponential. One finds that the function $\tilde Q$ has a power law tail 
\be \label{eq:powlaw}
\tilde Q( \tilde \Delta) \simeq C
  |\tilde \Delta|^{-(3+\frac{2}{s-1})}  \quad , \quad C =  \frac{\mathcal{N}}{\sqrt{\pi}} \frac{2^{\frac{s}{s-1}} \Gamma \left(\frac{3}{2}+\frac{1}{s-1}\right)}{c^{\frac{s}{s-1}} (s-1)}
\ee
where we used $\int_0^\infty du u^{\frac{3}{2} (s-1)} e^{- a u^{s-1}}= \frac{a^{\frac{1}{1-s}-\frac{3}{2}} \Gamma \left(\frac{3}{2}+\frac{1}{s-1}\right)}{s-1}$.
This power law tail is valid for $s>1$, and the tail exponent diverges as $s \to 1$. In summary we thus obtain for $s>1$
\begin{align}
 &
 Q_n(\Delta) \sim  \begin{cases} n^{1/2-\alpha} ~~~~~~~~~~~~~   \quad ,\quad  |\Delta| \ll n^{\alpha-1/2}\\
 n^{-\frac{s}{s+1}} |\Delta|^{-(3+\frac{2}{s-1})}  \quad ,\quad  n^{\alpha-1/2} \ll |\Delta|
 \end{cases} 
\end{align}
Note that the power law extends up to an upper cutoff of the order $\Delta_{\max} \sim 1/\sqrt{D(0)}$, which is determined by $D(0)$ (assuming here that
$D(h)$ is an increasing function of $h$).

For $0<s<1$, $\tilde h^{s-1} \Delta$ is a decreasing function of $\tilde h$, and one can evaluate \ref{eq:expQint} using a saddle point method. 
Upon the change of variable $\tilde h = v \tilde \Delta$ one can rewrite
\be 
\tilde Q( \tilde \Delta) =  \mathcal{N} \sqrt{\frac{c}{2\pi}}  |\tilde \Delta|^{\frac{3}{2} s- \frac{1}{2} }
\int_0^{+\infty} dv v^{\frac{3}{2} (s-1)} e^{- c \tilde |\Delta|^{s+1} f(v)} \quad , \quad f(v) = \frac{1}{2} \frac{1}{v^{1-s}} +\frac{2}{(s+1)^2} v^{s+1} 
\ee 
The function $f(v)$ has a single minimum at $v= v_c = \frac{1}{2} \sqrt{1-s^2}$, and one obtains at 
large $\tilde \Delta$ the saddle point estimate


\be \label{eq:strectchdessm1}
\tilde Q( \tilde \Delta) \simeq \mathcal{N} \frac{\sqrt{s+1}}{2^s (1-s^2)^{(1-s)/2}} \frac{1}{|\tilde \Delta|^{1-s}} e^{-\frac{2^{1-s} c}{(s+1)(1-s^2)^{(1-s)/2}} |\tilde \Delta|^{s+1}}
\ee
Hence for $0<s<1$ the decay is a stretched exponential with an exponent $1<s+1<2$. 

While the distribution of the increments of length 1 provides insights into the behavior of the random walk, it is necessary to also
compute the distribution of the increments of arbitrary small length $\ell \ll n$ to obtain the multiscaling properties,
to which we now turn.

\subsubsection{Distribution of the height differences } 

To compute the distribution of the height differences $h_{n+\ell}-h_n$ and their moments
we will again resort to the continuum model defined above, which becomes exact for large $n, \ell \gg 1$. 
The correspondence is $n \to x$ and $\ell \to y$. In section \ref{} we derive the propagator of the
continuum process with $D(h)=c |h|^{s-1}$, which is simple when considering the absolute value $|h(x)|$. 
Its expression reads
\be 
P(|h|,x||h_0|,0) =
\frac{2 c}{(s+1) x} \sqrt{|h_0|} |h|^{s-\frac{1}{2}} e^{-\frac{2 c
   \left(|h|^{s+1}+|h_0|^{s+1}\right)}{(s+1)^2 x}}
   I_{-\frac{1}{s+1}}\left(\frac{4 c
   |h|^{\frac{s+1}{2}} |h_0|^{\frac{s+1}{2}}}{(s+1)^2
   x}\right)
\label{eq:propagatorAbsh}
\ee 
which is the probability that the absolute value of the height $h(z)$ is $|h|$ at $z=x$, given that it is $|h_0|$ at $z=0$.
It is normalized to unity when integrating $|h|$ from $0$ to $+\infty[$.
One can check that $P(|h|,x||h_0|,0)$ has a limit when $h_0 \to 0$ which coincides
exactly with the one-point PDF of $|h|$ given in \eqref{PDFhcont}. 

This setting is convenient to compute the probability of the increment of the absolute value of the height, $\Delta= |h(x+y)|-|h(x)|$, which
we do below. We consider the regime $y \ll x$, in which case, as we argue below, $h(x+y)$ and $h(x)$ typically share the same sign, such that the increments of $|h|$ and of $h$ have similar distributions.
Using the Markov
property of the process, and choosing the initial condition $h_0=0$, we can write
\be
\begin{split} \label{eq:expprobdiff}
& P_{x,y}(\Delta)  = \int_{\max(0,-\Delta)}^{+\infty} dh P(h+\Delta,x+y|h,x)  P(h,x|0,0) = \int_{\max(0,-\Delta)}^{+\infty} dh P(h+\Delta,y|h,0)  P(h,x|0,0)  \\
&= \frac{2\mathcal{N}}{s+1} \frac{c}{x^{\frac{s}{s+1}}y} \int_{\max(0,-\Delta)}^{+\infty} dh h^{s-1/2} (h+\Delta)^{s-1/2}  e^{-\frac{2 c
   \left((h+\Delta)^{s+1}+h^{s+1}\right)}{(s+1)^2 y}}
   I_{-\frac{1}{s+1}}\left(\frac{4 c
   (h+\Delta)^{\frac{s+1}{2}} h^{\frac{s+1}{2}}}{(s+1)^2
   y}\right) e^{-\frac{2 c
   h^{s+1}}{(s+1)^2 x}}
\end{split}
\ee
We recall the asymptotics, 
$I_\nu(z) \simeq e^z/(\sqrt{2 \pi z})$ for $z \gg 1$, 
and $e^{- \frac{1}{2} (a^2 + b^2) } I_\nu(a b) \simeq e^{- \frac{1}{2} (a-b)^2}/\sqrt{2 \pi a b}$ 
for $a,b \gg 1$. 

 The contribution of the integral 
for $h \gg y^\alpha$ takes the form
\bea \label{eq:eqhgrand}
&&   P_{x,y}(\Delta)  \simeq \frac{\mathcal{N}}{\sqrt{2 \pi}} \frac{\sqrt{c}}{x^{\frac{s}{s+1}}\sqrt{y}} \int_{y^\alpha}^{+\infty} dh h^{3(s-1)/4} (h+\Delta)^{3(s-1)/4}  e^{-\frac{2 c
   \left((h+\Delta)^{(s+1)/2}-h^{(s+1)/2}\right)^2}{(s+1)^2 y}} e^{-\frac{2 c h^{s+1}}{(s+1)^2 x}}
\eea 
Consider now $|\Delta| \ll y^\alpha$, one can expand in $\Delta$ and this contribution becomes
\be
 P_{x,y}(\Delta)  \simeq \frac{\mathcal{N}}{\sqrt{2 \pi}} \frac{\sqrt{c}}{x^{\frac{s}{s+1}}\sqrt{y}}  \int_{y^\alpha}^{+\infty} dh h^{3(s-1)/2}   e^{-\frac{c 
   \Delta^{2} h^{s-1}}{2 y}} e^{-\frac{2 c 
   h^{s+1}}{(s+1)^2 x}}
\ee
which, for $y=1$ and $x=n$ is exactly the result in \eqref{eq:exprQdis}, obtained for the discrete model. We verify below that this indeed represents the relevant contribution for $|\Delta| \ll y^\alpha$. This result indicates that for $u \in [x, x+y]$ with $x \ll y$, the function $h(u)$ typically maintains the same sign. Consequently, the distribution of the increments of $h$ and of $|h|$ are similar. Hence, for $|\Delta| \ll y^\alpha$, the scaling function is given by
\be \label{eq:scalxy}
P_{x,y}(\Delta) \simeq x^{\frac{1}{2}-\alpha} y^{-\frac{1}{2}} \tilde Q\left( \Delta x^{\frac{1}{2}-\alpha} y^{-\frac{1}{2}} \right),
\ee
where the function $\tilde Q$ has been calculated in the previous section, see Eq. \eqref{eq:expQint}. 

To obtain the behavior for large $\Delta \gg y^\alpha$, we need to examine the contribution from small $h \ll y^\alpha$ to the integral in \eqref{eq:expprobdiff}. We focus first on $\Delta > 0$ for clarity. Since in that regime one has $h \ll \Delta$ one can expand the argument of the exponential
in small $h$ which gives an upper cutoff on $h$ of order $y/\Delta^s$. This implies that the argument of the Bessel function
is small, and we can use the asymptotic form $I_\nu(z) \sim \frac{z^\nu}{2^\nu \Gamma(1+\nu)}$ for $z \ll 1$.
Recalling that we consider here $y \ll x$, we obtain the contribution
\be
P_{x,y}(\Delta) \simeq \mathcal{N} \frac{(s+1)^{\frac{1-s}{s+1}}}{\Gamma\left(\frac{s}{s+1}\right)} \frac{(2 c)^{\frac{s}{s+1}}}{x^{\frac{s}{s+1}} y} \int_0^{y^\alpha} dh \, y^{\frac{1}{1+s}} h^{s-1} (h+\Delta)^{s-1} e^{-\frac{2 c \left((h+\Delta)^{s+1}+h^{s+1}\right)}{(s+1)^2 y}}.
\ee

First note that for $\Delta \ll y^\alpha$, this term is indeed negligible compared to \eqref{eq:scalxy},
Indeed, it is of order $\sim x^{-s/(s+1)} y^{(s-1)/(s+1)}$, which is precisely the same order
as is reached in \eqref{eq:scalxy} far in the tail for $\Delta \sim y^\alpha$ as
can be checked using the asymptotics \eqref{eq:powlaw} of the scaling function $\tilde Q$. 
This justifies
a posteriori the above claim that the integral in \eqref{eq:expprobdiff} is dominated by $h \gg y^\alpha$
when $\Delta \ll y^\alpha$. 

Returning to the case $\Delta \gg y^\alpha$, we see that in the region of integration, $h \ll y^\alpha \ll \Delta$, we can expand to first order in $h$. By making the change of variable $u = h \frac{2 c \Delta^s}{y (s+1)}$, we obtain
\be \label{eq:powlawstretchcutoff}
P_{x,y}(\Delta) \simeq \mathcal{N} \frac{(s+1)^{\frac{s^2+1}{s+1}} \Gamma(s)}{\Gamma\left(\frac{s}{s+1}\right) (2 c)^{\frac{s^2}{1+s}}} \frac{y^{\frac{s^2}{1+s}}}{x^{\frac{s}{1+s}} \Delta^{s^2+1-s}} e^{-\frac{2 c \Delta^{s+1}}{(s+1)^2 y}}.  
\ee
where we have used that $\int_0^{\frac{2 c y^{\alpha-1} \Delta^s}{(s+1)^2}} du \, u^{s-1} e^{-u} \simeq \Gamma(s)$
which holds since we consider the case $\Delta \gg y^\alpha$, $y^{\alpha-1} \Delta^s \gg 1$.
Note that, whether we consider the region of integration $h \le y^\alpha$ or $h \ge y^\alpha$, there is an exponential cutoff at $h_c \sim y/\Delta^s \ll y^\alpha$. Therefore, the first region is the relevant one. 

In summary we find that the PDF of the increments, $P_{x,y}(\Delta)$, takes schematically the form, for $y \ll x$ and $s>1$
\begin{align} \label{eq:distincrRW}
 &
 P_{x,y}(\Delta) \sim  \begin{cases} x^{\frac{1}{2}-\alpha} y^{-\frac{1}{2}}~~~~~~~~~~~~~~~~~   \quad ,\quad  |\Delta| \ll x^{\alpha-1/2} y^{1/2}\\
 x^{-\frac{s}{s+1}} y^{\frac{s}{s-1}} |\Delta|^{-(3+\frac{2}{s-1})}  \quad ,\quad  x^{\alpha-\frac{1}{2}} y^{-\frac{1}{2}} \ll |\Delta| \ll y^\alpha
 \end{cases} 
\end{align}
with a stretched exponential decay given by \eqref{eq:powlawstretchcutoff} for $\Delta$ larger than the cutoff $\sim y^\alpha$. 
Note that in the case $\Delta < 0$, we can see from Eq.\eqref{eq:expprobdiff} that making the change of variable $h' = h + \Delta$ leads to the same equation as for $\Delta > 0$, except that the last term is replaced by $e^{-\frac{2 c (h+|\Delta|)^{s+1}}{(s+1)^2 x}}$. 
However, as we assumed $y \ll x$, the dependence in $\Delta$ in this term turns out to be irrelevant in all the regimes that we considered.
This is presumably a signature that for $y \ll x$ there are essentially no sign changes of $h$. 

For $s<1$ there is no power law behavior for the increment distribution and one finds instead a stretched exponential form

\begin{align} \label{eq:distincRW2}
 &
 P_{x,y}(\Delta) \sim  \begin{cases} x^{\frac{1}{2}-\alpha} y^{-\frac{1}{2}}~~~~~~~~~~~~~~~~~~~~~~~~~~~~~~~~   \quad ,\quad  |\Delta| \ll x^{\alpha-1/2} y^{1/2}\\
 x^{\frac{-s(1-s)}{2(s+1)}} y^{-\frac{s}{2}}\frac{1}{|\Delta|^{1-s}} e^{-b_s \left|\frac{\Delta}{x^{\alpha-\frac{1}{2}} y^{\frac{1}{2}}}\right|^{s+1}}  \quad ,\quad  x^{\alpha-\frac{1}{2}} y^{-\frac{1}{2}} \ll |\Delta| 
 \end{cases} 
\end{align}
where the stretched exponential constant is given by $b_s=\frac{2^{1-s} c}{(s+1)(1-s^2)^{(1-s)/2}}$.

\subsubsection{Moments of the height differences } 

We can now compute the moments of order $q$ of the height increments, $\Delta= |h(x+y)|-|h(x)|$, from its distribution
in the case $y \ll x$. For $s>1$, there are two regimes in $q$.

In the first regime, for $q<q_c$, one can use \eqref{eq:scalxy} which leads to 
\begin{align}
 &
 \langle |\Delta|^q \rangle \simeq c_q y^{\frac{q}{2}} x^{q \, (\alpha-\frac{1}{2})}    \quad ,\quad  q < q_c =2+ \frac{2}{s-1} 
\end{align}
where $c_q= \int_{-\infty}^{+\infty} d\tilde \Delta |\tilde \Delta|^q \tilde Q(\tilde \Delta)$ is a $s$-dependent amplitude equal to
the $q$-th moment of the scaled distribution given in \eqref{eq:scalxy} and \eqref{eq:expQint}. We recall that $\alpha=1/(s+1)$.
Since this distribution
has a power law tail $\tilde Q(\tilde \Delta) \sim \tilde \Delta^{-(3+ 2/(s-1))}$, see \eqref{eq:powlaw},
this can only hold for $q< q_c=2+ \frac{2}{s-1}$, as $c_q$ diverges as $q \to q_c^-$.

For $q>q_c$ the moment is instead dominated by the cutoff of the distribution of $\Delta$, which has a fast stretched exponential  
decay, see \eqref{eq:powlawstretchcutoff} which leads to 

\begin{align}
 \langle |\Delta|^q \rangle \sim y^{q \alpha + \frac{s}{s+1}} x^{-\frac{s}{s+1}}  \quad ,\quad  q > q_c 
\end{align}

For $q=2$, note that, using \eqref{eq:scalxy}, one retrieves exactly \eqref{eq:scalemom2} obtained in the discrete setting. Recalling that the local roughness exponent is defined by $\langle |\Delta|^q \rangle^{1/q} \sim 
 y^{\alpha_{\rm loc}(q)} x^{\alpha - \alpha_{\rm loc}(q)}$, one thus finds for $s>1$

\begin{align}
 &
 \alpha_\text{loc}(q) \sim  \begin{cases} \frac{1}{2} ~~~~   \quad ,\quad  q < q_c \\
 \alpha + \frac{1}{q} \frac{s}{s+1}  \quad ,\quad  q_c < q.
 \end{cases} 
\end{align}
For $s \to 1$, one has $\alpha = \alpha_\text{loc}(q) =1/2$ and $q_c \to + \infty$ thus retrieving the standard EW scaling.

In the case $s<1$, since the distribution of increments has a fast stretched exponential decay, see
\eqref{eq:strectchdessm1}, 
there is a single regime in $q$, with no multiscaling, and one finds, using \eqref{eq:scalxy} and \eqref{eq:expQint}, that the moments behave as
\begin{align}
 &
 \langle |\Delta|^q \rangle \simeq c_q  y^{\frac{q}{2}} x^{q \, (\alpha-\frac{1}{2})}
\end{align}
leading to a local exponent $\alpha_\text{loc}(q)=1/2$ for $s<1$.



\section{Solution of the continuum random walk model}

The continuum model is defined by the Ito stochastic equation for $h(x)$
\be \label{1} 
dh(x) = \frac{1}{\sqrt{D(h(x))}}  dB(x)
\ee 
where $D(h)$ is a positive even function of $h$
with $D(0)$ finite. 

\subsection{Mapping to a Langevin equation}

One can transform this equation into one with additive noise.
We introduce the function ${\sf Y}(h) = \int_0^h dh \sqrt{D(h)}$.
We consider the process $Y(x)= {\sf Y}(h(x))$. 
Then using Ito rule, we obtain that it obeys the stochastic equation
\be 
d Y(x) = dB(x) + \frac{1}{4} \frac{D'(h(x))}{D(h(x))^{3/2}} dx  
\ee  
Inverting $Y={\sf Y}(h)$ as $h={\sf h}(Y)$, the process $Y(x)=Y(h(x))$ satisfies
now the following Langevin equation with additive noise and with a drift
\be \label{eq:LangevinUp}
dY = dB(x) - U'(Y) dx \quad , \quad - U'(Y)=  \frac{1}{4} \frac{D'({\sf h}(Y))}{D({\sf h}(Y))^{3/2}} 
\ee
which describes thermal motion (at temperature $1/2$) in the potential $U(Y)= - \frac{1}{4} \log D(h(Y))$. 
\\

Equivalently, one can perform the same transformation on the Fokker Planck equation. 
The PDF of $h(x)$ satisfies \eqref{FKPh}. Defining the PDF of $Y(x)$ as $Q(Y,x)$,
one obtains the relation $P(h,x) = \sqrt{D(h)} Q(Y,x) $. Inserting in \eqref{FKPh}
and using $\partial_h = Y'(h) \partial_Y = \sqrt{D(h)} \partial_Y$, one obtains
the Fokker Planck equation for $Q(Y,x)$ 
\be 
\partial_x Q = \partial_Y  (  \frac{1}{2}  \partial_Y Q + U'(Y) Q) 
\ee 
where $U'(Y)$ is given in \eqref{eq:LangevinUp}. This is indeed the FP associated
to the process \eqref{eq:LangevinUp}. 

Note that it admits formally a zero current equilibrium Gibbs measure
$Q_0(Y) \sim e^{-2 U(Y)} = \sqrt{D(Y)}$, i.e. $P_0(h) = D(h)$, which
however is normalizable only if $D(h)$ decays faster than $1/h$, i.e. $s<0$.,
which is not the regime considered here. 

\subsection{Power law $D(h)$ and mapping to a Bessel process}

Let us now specialize to the power law case $D(h) = c |h|^{s-1}$. In that case one has 
\be 
D(h) = c|h|^{s-1}   ~,~ 
{\sf Y}(h) = \sqrt{c} \frac{2}{s+1}   |h|^{\frac{s+1}{2}} {\rm sgn} h ~,~  
{\sf h}(Y) = (\frac{s+1}{2 \sqrt{c}} Y)^{2/(s+1)} 
{\rm sgn} Y
~,~  D(Y)  =  (c^{\frac{1}{s-1}} \frac{s+1}{2} |Y|)^{\frac{2(s-1)}{s+1}}
\ee 
The potential $U(Y)$ is then $U(Y)= -  \frac{s-1}{2(s+1)} \log |Y|$.
It is thus repulsive for $s>1$ and attractive towards $Y=0$ for $0<s<1$. 
In both cases however for $s>0$, $Y$ undergoes unbounded diffusion in the potential $U(Y)$,
and obeys the stochastic equation
\bea  
&& dY = dB(t) - U'(Y) = dB(t) + \frac{s-1}{2(s+1)} \frac{1}{Y}
\eea 
Thus one finds that up to its sign, $Y(x)$ is a Bessel process.
More precisely
\be 
Y(x) = ( {\rm sgn} Y(x) ) R(x) \quad , \quad |Y(x)|= R(x)
\ee 
Here $R(x)$ is a Bessel process \cite{WikipediaBesselProcess},
which is defined as the Euclidean norm of the $d$ dimensional
Brownian motion ${\bf W}(x)$. One has 
\be 
R(x) = ||{\bf W}(x)||  
\quad , \quad 
d=  \frac{2 s}{s+1} = 2 \nu + 2  \quad , \quad \nu = - \frac{1}{1+s} 
\ee 
where $W(x)$ is the $d$-dimensional BM. Thus one finds that $Y(x)$ is recurrent for any $s>0$, since $d<2$
and $||{\bf W}(x)|| $ crosses infinitely often the origin.

To use the properties of the Bessel process we will study $|Y(x)|$. 
The propagator of the Bessel process, $Q(|Y|,x||Y_0|,0)$, is known and reads for $\nu>-1$ (i.e. for $s>0$)
\be 
Q(|Y|,x||Y_0|,0) = \frac{|Y|^{\nu +1} Y_0^{-\nu } e^{\frac{-Y^2-Y_0^2}{2
   x}} I_{\nu }\left(\frac{|Y| Y_0}{x}\right)}{x} \quad , \quad \nu = - \frac{1}{1+s} 
\ee 
where $I_\nu$ is the modified Bessel function.
One can check that the current near $Y=0^+$ 
\be
j(Y,x) = - \partial_Y Q + \frac{2 \nu + 1}{2 Y} Q \propto
(2 (1+ \nu) x - Y_0^2) Y^{2 + 2\nu} 
\ee 
vanishes for $\nu > -1$, which corresponds to reflecting boundary conditions. 
\\


Finally, using that $|Y|= \sqrt{c} \frac{2}{s+1} |h|^{(s+1)/2}$, we readily obtain the propagator in
the $h$ variable, which is given in \eqref{eq:propagatorAbsh}. 

{\bf Remark}. Note that the above continuum model \eqref{1} was studied previously 
in \cite{fa2003power}, where the one-point distribution was obtained, as well
as very recently in \cite{del2024generalized}, 
although by different methods, and focusing on different observables, such as first passage and occupation times.
\\

{\bf Remark} The present model is also related to the ABBM model of avalanches, 
see e.g. Eq (213) in \cite{le2009driven}. The identification is 
thus $u=x$ and
\be 
{\sf v}(u) \equiv Y(x) \quad , \quad m^2 v = \frac{d-1}{2} = \frac{s-1}{2(s+1)} 
\quad , \quad \sigma=1/2
\ee
Our present RW model corresponds to the limit $m \to 0$. The avalanche exponent (231) for the ABBM model is
\be 
\tau = 2 - \frac{d}{2} = \frac{3}{2} - \frac{m^2 V}{2 \sigma} = \frac{2 + s}{1 + s} 
\ee 
Here it describes the distribution of the $x$ where the process $h(x)$ 
returns to zero (up to some small cutoff, see discussion in (232) there).  

\section{Numerical results}


\subsection{Numerical details} 


We use an Euler integration scheme to integrate Eq.\eqref{eq:SPMEdx}, namely
\be   
\label{discretespme} 
h_n(t + \Delta t) = h_n(t) +  \Delta t \left[ D(h_n(t))(h_{n+1}(t)-h_n(t)) + D(h_{n-1}(t))(h_{n-1}(t) - h_n(t))  \right] + 2 \sqrt{\Delta t} R_n(t),  
\ee
where $R_n(t)$ is a Gaussian random variable with unit variance, and $\Delta t$ is the integration step. To ensure numerical stability, the integration step $\Delta t \, |D(h_n)(h_{n+1}-h_n)+D(h_{n-1})(h_{n-1}-h_n)| $ should be small compared to $h_n$. This requires a small $\Delta t$ when $s>1$, $h$ large.
Knowing that $h \sim t^{\beta} = t^{1/(3+s)}$ and $D(h) \sim h^{s-1}$, we therefore take for $s>1$ 
\be
\Delta t = 0.02 \max{(1,t)^{\frac{s-1}{3+s}}}.
\ee
For $s<1$, we set $\Delta t=0.02$.

One possible choice for $D(h)$ is 
\be \label{formDh}
D(h) =   \left(\nu^{1/(s-1)}+ |h|\right)^{s-1}
\ee
with $s > 0$. This choice preserves the parity of $D(h)$ and regularizes its behavior at the origin with $D(0) = \nu$. While $D(h)$ is non-analytic at $h = 0$, this poses no numerical issues. In particular, we verified that it satisfies the scaling relations from the RG.




In order to reach a stationary state, we must consider a situation where the center of mass of the interface remains confined. This is not the case under periodic boundary conditions, as the center of mass undergoes diffusion. Instead, we consider the case where one extremity of the interface is fixed (i.e., \( h_0 = 0 \)) while the other extremity, \( h_L \), is free (i.e., \( D(h_L) = 0 \)). In this setup, the center of mass of the interface will scale as \( \sim L^\alpha \) at large times.



\subsection{Numerical evidence for the correspondence between stationary interface and the random walk model}

In the following, we study the correspondence between a stationary interface, with the right boundary fixed at $h_0$ and the left boundary free, and the random walk model defined by:

\be
\tilde{h}_{n+1} = \tilde{h}_n + \frac{1}{\sqrt{D(\tilde{h}_n)}} \xi_n.
\ee

As mentioned in the main text, we expect this correspondence to hold when the interface varies slowly, i.e.
\be\label{eq:condgauss1SM}
|h_{n+1} - h_n| \ll \frac{D(h_n)}{|D'(h_n)|}.
\ee
When this condition is met, $|h_{n+1} - h_n| \sim 1/\sqrt{D(h_n)}$ and, considering the chosen form \eqref{formDh} of $D(h)$, this leads to the condition:

\be
\left| \frac{D(h_n)^{3/2}}{D'(h_n)} \right| \sim \left(\nu^{1/(s-1)} +  |h_n|\right)^{(s+1)/2} \gg 1.
\ee

This condition is always satisfied for sufficiently large values of $h_n$. However, it can also hold for small $h_n$ if $\nu$ is large enough for $s > 1$, or small enough for $s < 1$.

In Fig.~\ref{fig:compSPMERWs05} (Left),  we test the validity of the condition \eqref{eq:condgauss1SM}. For small increments, the distribution of $h_{n+1}$ conditioned to $h_n$ becomes Gaussian of mean $h_n$ and variance $1/D(h_n)$. For large increments, deviations from the Gaussian behavior are observed. However,
these deviations disappear when $|h_n| \gg 1$, in which case the condition \eqref{eq:condgauss1SM} is satisfied.
Numerically, the distribution of $h_{n+1}$ is obtained by considering a small window around $h_n$. In practice, we take $\left[-0.07,0.07\right]$ for $h_n=0$; $\left[0.45,0.55\right]$ for $h_n=0.5$; $\left[1.9,0.2.1\right]$ for $h_n=2$; $\left[3.8,4.2\right]$ for $h_n=4$.

In Fig.~\ref{fig:compSPMERWs05} (Right), we show the distribution of $h_n$ (with $n \sim L$) for an interface with the left boundary fixed at $h_0 = 300$ for $s=0.5$. We choose $h_0 \gg 1$ and $L$ sufficiently small such that $h_L \gg 1$. Under these conditions, we expect the SPME interfaces to be well-described by the random walk model. The prediction from the continuous RW \eqref{eq:propagatorAbsh}) is for the process $|h|$, but which is equivalent to the process $h$ here as $h_n$ remains positive.
The random walk model provides a good approximation, although some discrepancies appear for smaller values of $h_n \sim 150$.


\begin{figure} 
    \includegraphics[width=0.48\textwidth]{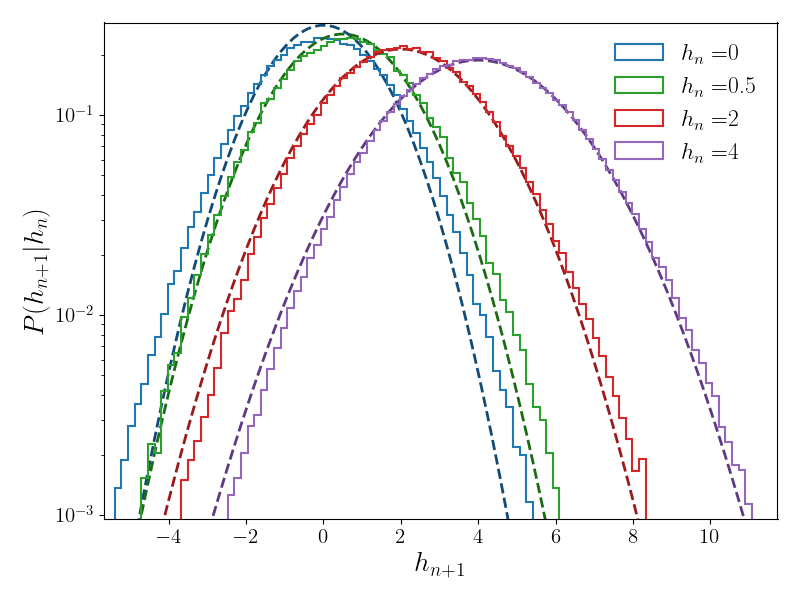}
    \includegraphics[width=0.48\textwidth]{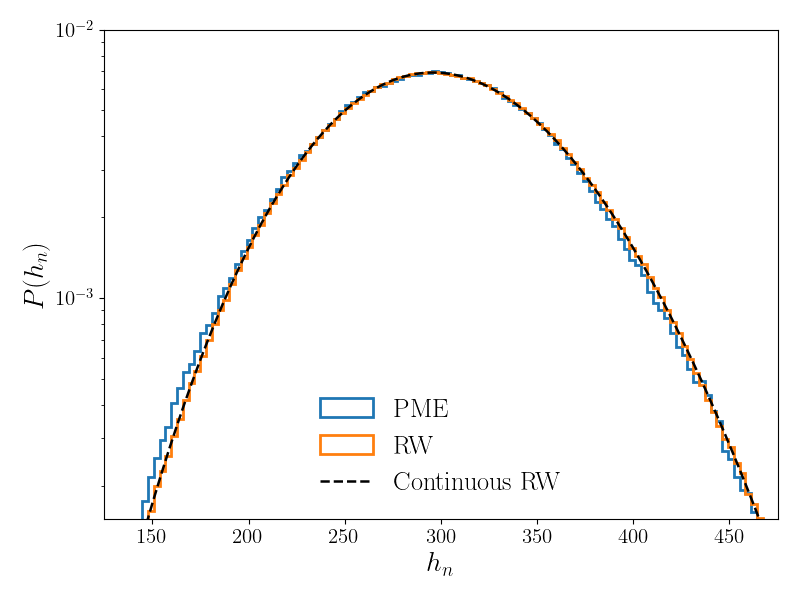}
    \caption{Left: Distribution of the height $h_{n+1}$ conditioned to a given $h_n$ for stationary interfaces of the SPME for $D(h)=0.5/\sqrt{1+|h|}$ ($s=0.5$) (Histogram). The prediction for the random walk defined in Eq.\eqref{eq:RW} is a Gaussian of mean $h_n$ and variance $1/D(h_n)$ (Continuous lines) Right: PDF of $h_n$ averaged over $n \in [L-p,L]$ with $L=100$ and $p=10$, with boundary condition $h_0=300$
and $D(h)=0.5/\sqrt{1+|h|}$ ($s=0.5$) for (i) the stationary SPME (blue), (ii) the discrete RW model (orange),
(iii) the analytical prediction (black) from the continuum diffusion model (using the
two point transition probability in 
\eqref{eq:propagatorAbsh}).}
    \label{fig:compSPMERWs05}
\end{figure}

\begin{figure} 
    \includegraphics[width=0.55\textwidth]{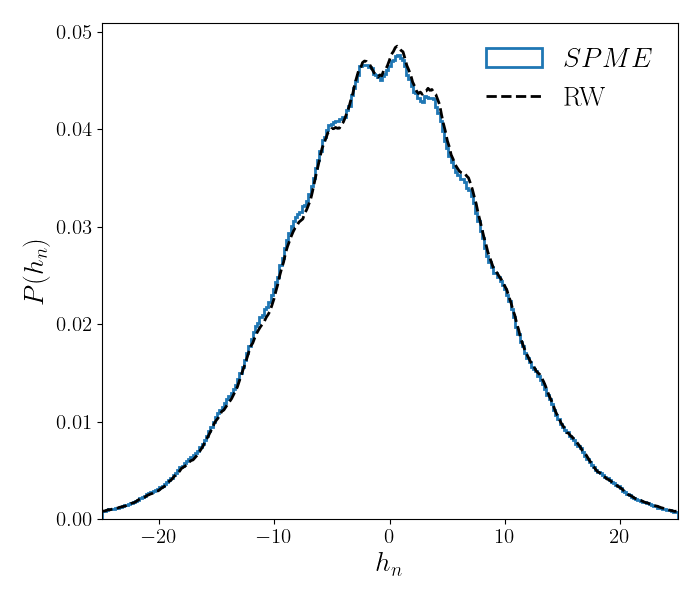}
    
    \caption{PDF of $h_n$ for the SPME averaged over $n \in [L-p,L]$ with $L=100$ and $p=50$, with boundary condition $h_0=0$ and $D(h)=1 + \sin{(2 h)}/10$. Compared to the random walk (black discontinuous line), obtained numerically.}
    \label{fig:compRWDh}
\end{figure}

The SPME/RW mapping is expected to be satisfied for any function that varies slowly, i.e., satisfies the condition \eqref{eq:condgauss1SM}. As an example, in Fig.~\ref{fig:compRWDh}, we show the distribution of $h_n$ for an interface starting from $h_0=0$ and with $D(h) = 1 + \sin{(2h)}/10$, and observe a good agreement with the random walk obtained numerically. We use the discretization of the Laplacian $\partial_{xx} V(h) \to V(h_{n+1}) + V(h_{n-1}) - 2 V(h_{n})$ with $V(h)=h - \cos{(2h)}/20$.


We have shown that when the condition \eqref{eq:condgauss1SM} is satisfied, the mapping between the RW and the stationary SPME is excellent. Now we study cases where the condition \eqref{eq:condgauss1SM} only in the asymptotic limit and deviations are expected when the interface is close to the origin. In particular, this is the case flr the rescaled height $h_L/L^\alpha$ for an interface starting at the origin $h_0=0$. In Fig. \ref{fig:histPMh}, we show that the rescaled height differ from the prediction of the RW, both for $s<1$ and $s>1$. The behavior is qualitatively correct, but even at $L=400$ and $L=1000$, we are still far from the asymptotic prediction, presumably because $h \sim L^\alpha$ is not large enough (in particular for $s=3$, since $\alpha = 0.25$).





\begin{figure}
    \centering
    \includegraphics[width=0.45\textwidth]{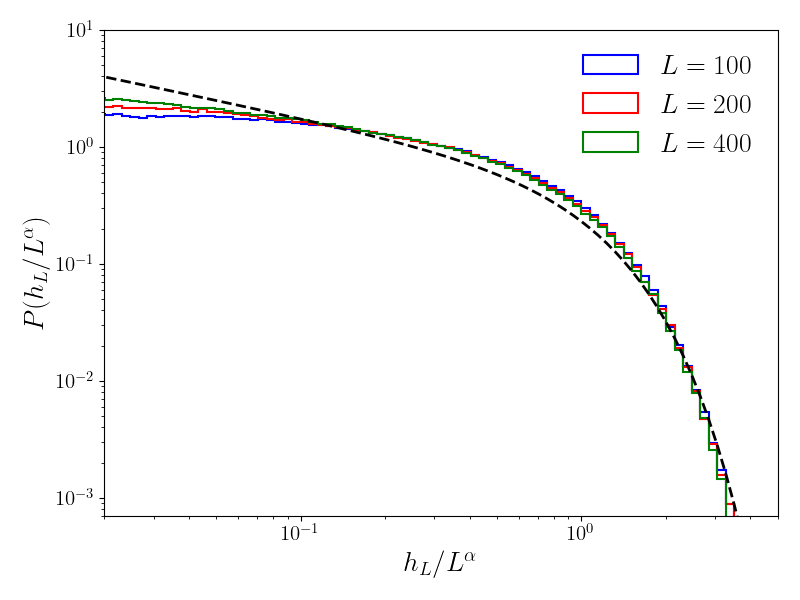}
    \includegraphics[width=0.45\textwidth]{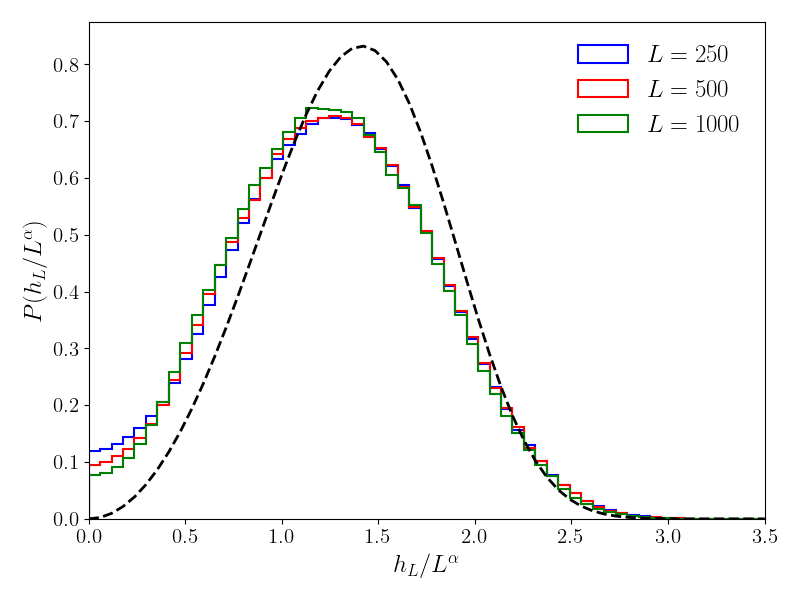}
    \caption{PDF of $h_L/L^\alpha$, with boundary condition $h_0=0$ and several $L$. The predicted (RW) asymptotic behavior is plotted in black dotted line. (Left) $D(h)=0.5/\sqrt{1+|h|}$ ($s=0.5$) (Right) $D(h)=1+h^2$ ($s=3$). The PDF is averaged over $n \in [L-p,L]$ with $p=10$.}
    \label{fig:histPMh}
\end{figure}

In Fig. \ref{fig:colldistdeltah}, we also study the distribution of the height differences $\Delta = h_{n+\ell}-h_n$ for $s>1$ for which the multifractal behavior of $\alpha_{\rm loc}(q)$ is expected. We correctly recover the behavior of the typical height difference $\Delta_{\rm typ} \sim L^{\alpha - \frac{1}{2}} \ell^{1/2}$ and its cutoff $\Delta_c \sim \ell^\alpha$. The power law decay, predicted in \eqref{eq:distincrRW}, is underestimated as we find $\sim 1/\Delta^3$ instead of $1/\Delta^4$. Larger interfaces are needed to reach the asymptotic limit.





\begin{figure}
    \centering
    \includegraphics[width=0.55\textwidth]{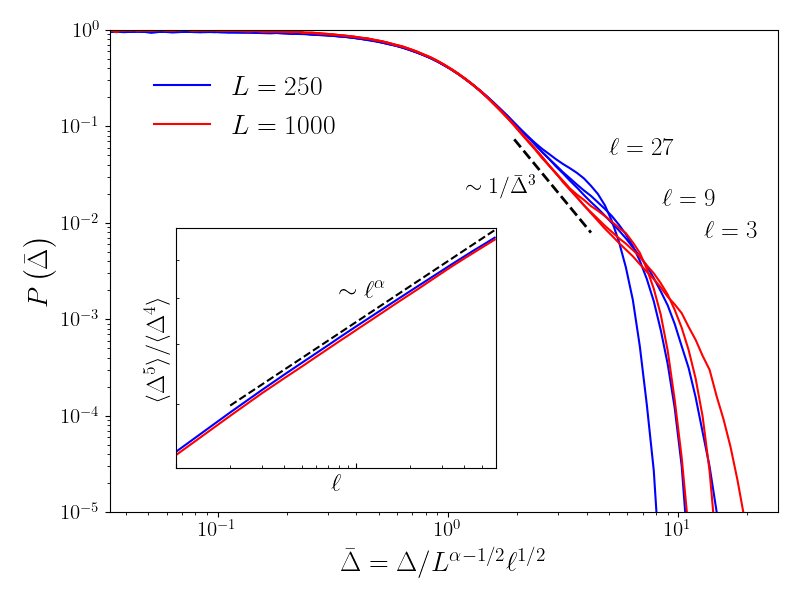}
    \caption{Main panel: Distribution of $P(\bar \Delta)$,  $\bar \Delta= \Delta/( L^{\alpha - \frac{1}{2}} \ell^{\frac{1}{2}})$ for $s=3$, $\alpha=1/4$ (blue line: $L=250$, red line: $L=1000$) and for three different values of $\ell$. We have taken $n \in [L-100,L]$. The collapse at small $\bar \Delta$ shows the scaling $\Delta_\text{typ}(\ell) \sim \sqrt{\ell} L^{\alpha -1/2}$. The plot also shows a power law decay with exponent $3$ for $L=1000$. Inset: Determination of $\Delta_\text{max}(\ell) \sim \langle \Delta^5 \rangle/\langle \Delta^4 \rangle$.}
    \label{fig:colldistdeltah}
\end{figure}

\subsection{Structure Factor}

The structure factor $S_q(t)$ is a crucial measure that probes the spatial correlations of interface heights. It is defined as:

\begin{equation}
S_k(t) = \left\langle\frac{1}{L} \left| \sum_{n=1}^{L} h_n(t) e^{-iqn} \right|^2\right\rangle = \langle \hat{h}_k(t) \hat{h}_{-k}(t) \rangle,
\end{equation}

where $\hat{h}_k$ denotes the Fourier transform of the height profile $h_n$. The standard scaling behavior in the stationary regime is $S_k \sim k^{-(d+2\alpha)}$ where $d$ is the spatial dimension and $\alpha$ the roughness exponent. However, from the mapping to the random walk, we expect the SPME to display anomalous scaling properties. This leads to the prediction, for $1/L < k \ll 1$,

\begin{equation}
S_k \sim L^{ 2 (\alpha - \alpha_{\rm loc}(2))}/k^{1 + 2 \alpha_{\rm loc}(2)},
\end{equation}

with $\alpha_{\rm loc}(2)=1/2$ for all $s$ since $q_c= 2+\frac{2}{s-1}>2$. We thus expect:

\begin{equation}
S_k \sim L^{ 2 \alpha - 1}/k^{2},
\end{equation}

for $1/L < k \ll 1$. This can be seen in Fig. \ref{fig:collSq}. For $s=0.5$ (Left), this scaling is satisfied. For $s=3$, the collapse for several $L$ improves as $L$ increases and the slope is $\sim 1/k^{1.9}$, slightly smaller than the predicted $\sim 1/k^2$. This is likely due to the finite size effects already observed in the distribution of the jumps in Fig. \ref{fig:colldistdeltah}. 

\begin{figure}
    \centering
    \includegraphics[width=0.47\textwidth]{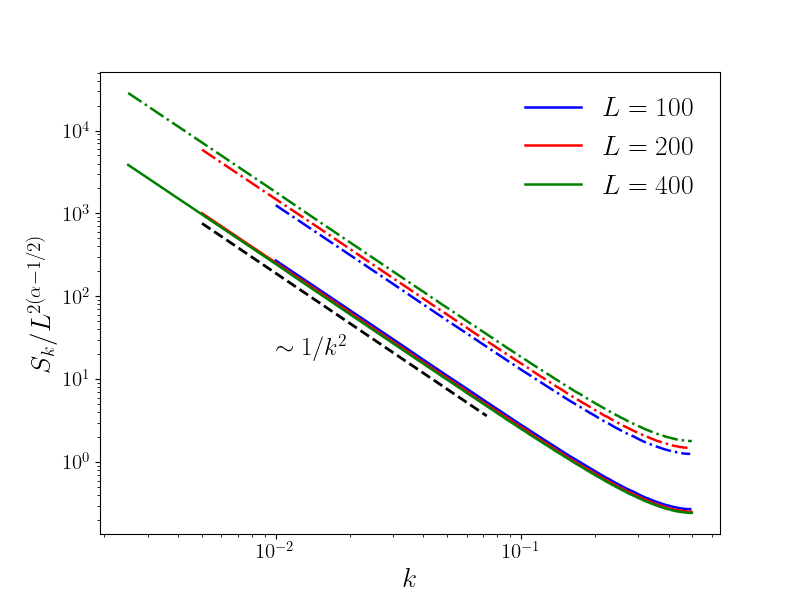}
    \includegraphics[width=0.47\textwidth]{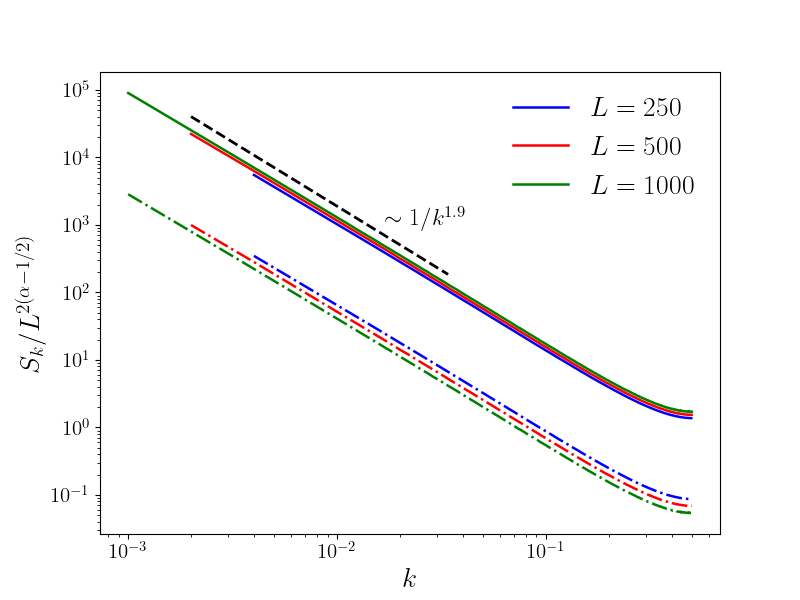}
    \caption{Structure factor $S_k$ for (Left) $D(h) = 1/(1 + |h|)$ corresponding to $s=0.5$, $\alpha=2/3$ and (Right) $D(h) = 1/4 + h^2$, corresponding to $s=3$, $\alpha=1/4$. The continuous line corresponds to the collapse $S_k/L ^{2(\alpha-1/2)}$ while the dotted line to $S_k$.}
    \label{fig:collSq}
\end{figure}

\newpage

\section{Fokker-Planck equation for the discrete PME}

Consider the discrete SPME equation, for $n=1,\dots,L$
\be
\partial_t h_n =  D(h_n) \left(h_{n+1}-h_{n} \right) + D(h_{n-1}) \left(h_{n-1}-h_{n} \right)  +\eta_n(t).
\ee
and recall that we choose $\eta_n(t)$ to i.i.d white noises with $\langle \eta_n(t) \eta_n(t') = 2 \delta_{nn'} \delta(t-t') $.
We focus here on the case where $h_0$ is fixed and $h_{L}$ free, which amounts to set $D(h_L)=0$. 

Denoting $\vec h = (h_1,\dots,h_L)$ the probability density function ${\cal P}(\vec h,t)$ evolves according
to the Fokker-Planck equation
\bea 
&& \partial_t \mathcal{P}(\vec h,t) = -\sum_{n=1}^L \frac{\partial}{\partial h_n} J_n(\vec h) \\
&& J_n(\vec h,t) = \left(D(h_n) \left(h_{n+1}-h_n \right) + D(h_{n-1}) \left(h_{n-1}-h_n\right)\right) \mathcal{P}(\vec h,t)  -  \frac{\partial}{\partial h_n} \mathcal{P}(\vec h,t)
\eea
where $J_n(\vec h,t) $ is the current. The current vanishes exactly when detailed balance is obeyed, which is not the case here
since we are dealing with a non equilibrium system.

With the above boundary conditions, our trial ansatz for the stationary measure 
corresponds to the probability distribution of $\vec h$ for the RW model 
introduced in the text \eqref{RWmodelSM}. We recall that it is defined by the recursion
\be \label{RWrecursion} 
h_{n+1}=h_n + \frac{1}{\sqrt{D(h_n)}} \xi_n  
\ee
This leads to the probability distribution of $\vec h$. 
It reads $
P_{RW}(\vec h) = \mathcal{N} e^{-F(\vec h)}$
 with 
\be
\label{eq:newansatz}
F(\vec h)=\sum_{n=0}^{L-1} [ D(h_n) \frac{(h_{n+1}-h_n)^2}{2} - \frac{1}{2} \log{D(h_n)}].
\ee

We will thus take this trial probability as our initial condition, and see how it evolves.
Hence we consider the evolution of $\mathcal{P}(\vec h,t)$ starting from
\be 
\label{eq:EqPinit}
\mathcal{P}(\vec h,t=0) = P_{RW}(\vec h)
\ee 
The current at time $t=0$ is then, for $1 \leq n \leq L-1$
\be
J_n(\vec h,t=0) = G_n(\vec h) P_{RW}(\vec h)  \quad , \quad G_n(\vec h) = D'(h_n) \frac{(h_{n+1}-h_n)^2}{2} - \frac{1}{2} \frac{D'(h_n)}{D(h_n)}
\ee
with $J_L(\vec h,t=0)=0$ and $G_L(\vec h) =0$.
One can check that $\sum_{n=1}^L \frac{\partial}{\partial h_n} J_n(\vec h,t=0)$ does not vanish exactly for finite $L$, 
hence the trial ansatz is not an exact stationary measure for finite $L$. However we claim that 
it becomes an increasingly good approximation of the true stationary measure of the SPME
in the large $L$ limit upon some appropriate rescaling. This is strongly supported by our numerical
simulations and we now provide a rationale for this claim.

It turns out that one can compute exactly the time derivative (at time $t=0$) of certain observables
with the initial condition given by the trial stationary measure. 

Consider an observable ${\cal O}(\vec h)$ and define ${\cal D} \vec h= \prod_{n=1}^L dh_n $. One has, for general
initial condition
\be  
 \frac{d}{d t } \int {\cal D} \vec h {\cal O}(\vec h) P(\vec h,t) 
= \int {\cal D} \vec h {\cal O}(\vec h) \frac{d}{d t }  P(\vec h,t) 
= - \int {\cal D} \vec h \sum_{n=1}^L {\cal O}(\vec h) 
 \frac{\partial}{\partial h_n} J_n(\vec h,t)  = \int {\cal D} \vec h \sum_{n=1}^L \frac{\partial {\cal O}(\vec h)}{\partial h_n} J_n(\vec h ,t)
\ee 
Applying this at $t=0$ with the initial condition \eqref{eq:EqPinit} 
\be \label{eq:formdtStatConj} 
\frac{d}{d t } \int {\cal D} \vec h {\cal O}(\vec h) P(\vec h,t)|_{t=0}  
= \int {\cal D} \vec h \sum_{n=1}^L \frac{\partial {\cal O}(\vec h)}{\partial h_n} G_n(\vec h) P_{\rm RW}(\vec h)  
=  \langle \sum_{n=1}^L \frac{\partial {\cal O}(\vec h)}{\partial h_n} G_n(\vec h) \rangle_{\rm RW} 
\ee 
where here we denote by $\langle \dots \rangle_{\rm RW}$ denote an expectation value with respect to our trial measure \eqref{eq:newansatz}. 
Hence \eqref{eq:formdtStatConj} gives a formula for the time derivative at time zero of observables when the system
starts at time zero with the trial stationary measure.

Let us consider the observable  ${\cal O}(\vec h)=  (h_{m+1}-h_m)^2 D(h_m) $ where $1 \leq m \leq L-1$ is given. 
For the RW model, and hence at $t=0$ here, from \eqref{RWrecursion} we see that
it corresponds to the increment ${\cal O}(\vec h)= \xi_m^2$, which is of order $1$ and independent of $h_j$ for $j \le m$. 
One can thus, at $t=0$, rewrite:
\be \label{eq:GRW}
2 G_n(\vec h) = \frac{D'(h_n)}{D(h_n)} ( D(h_n) (h_{n+1}-h_n)^2 - 1) = \frac{D'(h_n)}{D(h_n)}  (\xi_n^2 - 1),
\ee 
One also has:
\bea 
\partial_{h_n} {\cal O}(\vec h) = \delta_{n m} \left( (h_m-h_{m+1}) 2 D(h_m) + (h_{m+1}-h_m)^2 D'(h_m) \right) 
+ \delta_{n,m+1}  (h_{m+1}-h_m) 2 D(h_m) 
\eea 
One obtains the sum in \eqref{eq:formdtStatConj} as a sum of two terms
\be 
\langle \sum_{n=1}^L \frac{\partial {\cal O}(\vec h)}{\partial h_n} G_n(\vec h) \rangle_{\rm RW} = A + B 
\ee 
The first one is 
\bea
&& A = \langle G_m(\vec h) \left( (h_m-h_{m+1}) 2 D(h_m) + (h_{m+1}-h_m)^2 D'(h_m) \right) \rangle_{\rm RW} \\
&& = \frac{1}{2} \langle \frac{D'(h_m)}{D(h_m)}  (\xi_m^2 - 1) ( - \frac{\xi_m}{\sqrt{D(h_m)}}  2 D(h_m) 
+ \xi_m^2 \frac{D'(h_m)}{D(h_m)} ) \rangle_{\rm RW}  =  \langle \frac{D'(h_m)^2}{D(h_m)^2} \rangle_{\rm RW} 
\eea 
where we have used \eqref{RWrecursion}, \eqref{eq:GRW}, since we are computing expectation values at time $t=0$ with respect to
the random walk. We also used that $\xi_m$ is independent of $h_m$ and is a standard centered Gaussian random variable,
hence $\langle \xi_m^2 (\xi_m^2 - 1) \rangle_{\rm RW} = 2$.
The second one is 
\bea 
B = \langle G_{m+1}(\vec h) (h_{m+1}-h_m) 2 D(h_m)  \rangle_{\rm RW} = \frac{1}{2}
\langle \frac{D'(h_{m+1})}{D(h_{m+1})}  (\xi_{m+1}^2 - 1) \frac{\xi_m}{\sqrt{D(h_m)}} 2 D(h_m) \rangle_{\rm RW} = 0
\eea 
since $\xi_{m+1}$ is independent of the other variables and $\langle (\xi_{m+1}^2 - 1) \rangle_{\rm RW}=0$. 

In total we have shown that, starting from initial condition $P(\vec h,t=0)=P_{RW}(\vec h)$ one has 
\be \label{dt} 
\frac{d}{dt} \langle {\cal O}(\vec h) \rangle|_{t=0} = \langle \frac{D'(h_m)^2}{D(h_m)^2} \rangle_{\rm RW} 
\ee 
This is not zero, as it would be if the trial distribution was the exact stationary measure.

However we now show that the r.h.s of \eqref{dt}  vanishes at large $L$ as a power of $L$. To determine 
this power let us choose
$m$ of order $L$ and consider the 
the probability distribution $P_L(h)$ of $h=h_m$. We need to evaluate
\be 
\langle \frac{D'(h_m)^2}{D(h_m)^2} \rangle_{\rm RW} = \int  dh P_L(h) \frac{D'(h)^2}{D(h)^2}
\ee 
To take into account that there is a cutoff region at $h =O(1)$, with a finite $D(0)=O(1)$ we can use as an example the form
$D'(h)/D(h) \approx (s-1)/(1 + |h|)$. As we discussed in Section \ref{secPredRW} for $h \sim L^\alpha$
the distribution $P_L(h)$ takes the scaling form $P_L(h) \simeq L^{-\alpha} \tilde P(h/L^\alpha)$, 
which furthermore displays a power law behavior $P_L(h) \sim L^{\alpha-1} |h|^{s-1}$ 
for $1 \ll h \ll L^\alpha$. Hence the above integral contains two separate regions, and there are two cases.

(i) for $s>2$ the integral is dominated by the region $h \sim L^\alpha$ and the integral behaves as
\be 
\frac{d}{dt} \langle {\cal O}(\vec h) \rangle|_{t=0} \sim (s-1)^2  \int \frac{dh}{(1 + |h|)^2} L^{-\alpha} \tilde P(h/L^\alpha) \sim L^{-2 \alpha}.
\ee

(ii) for $s<2$ the integral is dominated by the region $h \ll  L^\alpha$ and the integral behaves as
\be 
\frac{d}{dt} \langle {\cal O}(\vec h) \rangle|_{t=0} \sim (s-1)^2 \int \frac{dh}{(1 + |h|)^2}  P(h) \sim \int dh P(h) \sim L^{-(1-\alpha)}.
\ee 

At the transition between the two cases i.e. for $s=2$ the decay is $L^{-2/3}$.

In summary we have shown that, starting from the random walk probability measure $P_{\rm RW}(\vec h)$ 
there is an observable ${\cal O}(\vec h)=  (h_{m+1}-h_m)^2 D(h_m) $
which 
is of order $O(1)$ at $t=0$ and whose time derivative vanishes at time zero in the limit
of a large system $L \to +\infty$. This is a strong indication that the random walk measure
describes well the stationary state of the SPME. Of course it is not a proof, unless we
could show that a similar result holds for a much larger class of observable.


{\bf Remark}. For some observables the time derivative at time zero vanishes exactly. Indeed, let us 
return to \eqref{eq:formdtStatConj} for a general observable. It reads
\bea
 \frac{d}{dt} \langle {\cal O}[\vec h] \rangle|_{t=0} = \frac{1}{2}  \langle \sum_{n=1}^L (\partial_{h_n} {\cal O}[\vec h]) \frac{D'(h_n)}{D(h_n)}  (\xi_n^2 - 1) \rangle
\eea 
An immediate consequence is that the time derivative (at time zero) of any function of the height at one point, 
${\cal O}[\vec h] = f(h_j)$ for a given $j$, vanishes since 
\be 
\frac{d}{dt} \langle f(h_j) \rangle|_{t=0} = \frac{1}{2}  \langle f'(h_j) \frac{D'(h_j)}{D(h_j)}  (\xi_j^2 - 1) \rangle = 0 
\ee 
Hence the one point marginal distribution of $h_j$ is unchanged in an infinitesimal time interval starting from $P_{RW}(\vec h)$. 

\end{widetext} 
\end{document}